\documentclass[pdftex,twocolumn,epjc3]{svjour3}          

\RequirePackage[T1]{fontenc}

\smartqed  

\RequirePackage{graphicx}
\RequirePackage{newtxtext,newtxmath}
\RequirePackage{flushend}
\RequirePackage[numbers,sort&compress]{natbib}
\RequirePackage[colorlinks,citecolor=blue,urlcolor=blue,linkcolor=blue]{hyperref}
\RequirePackage{amsmath}
\RequirePackage{doi}
\RequirePackage{color}

\RequirePackage[switch]{lineno}

\journalname{Eur. Phys. J. C}

\definecolor{darkgreen}{rgb}{0.0,0.4,0.0}
\DeclareMathAlphabet\mathbfcal{OMS}{cmsy}{b}{n}

\begin{document}

\title{Identification and simulation of surface alpha events on passivated surfaces of germanium detectors and the influence of metalisation}


\author{I.~Abt\thanksref{addr1}\and
    C.~Gooch\thanksref{addr1}\and
    F.~Hagemann\thanksref{addr1}\and
    L.~Hauertmann\thanksref{e1,addr1}\and
    X.~Liu\thanksref{addr1}\and
    O.~Schulz\thanksref{addr1}\and
    M.~Schuster\thanksref{addr1}\and
    A.~J.~Zsigmond\thanksref{addr1}
}

\thankstext{e1}{e-mail: lhauert@mpp.mpg.de (corresponding author)}

\institute{Max Planck Institut für Physik, Föhringer Ring 6, 80805 Munich, Germany\label{addr1}
}

\date{Received: date / Accepted: date}

\maketitle

\begin{abstract}
Events from alpha interactions on the surfaces of germanium detectors are a major contribution to the 
background in germanium-based searches for neutrinoless double-beta decay.
Surface events are subject to charge trapping, affecting their pulse shape and reconstructed energy.
A study of alpha events on the passivated end-plate of a segmented true-coaxial n-type high-purity germanium detector is presented.
Charge trapping is analysed in detail and an existing pulse-shape analysis technique 
to identify alpha events is verified with mirror pulses observed in the non-collecting channels of the segmented test detector.
The observed radial dependence of charge trapping confirms previous results.
A dependence of the probability of charge trapping on the crystal axes is observed for the first time.
A first model to describe charge trapping effects within the framework of the simulation software \emph{SolidStateDetectors.jl} is introduced.
The influence of metalisation on events from low-energy gamma interactions close to the passivated surface is also presented.
\end{abstract}

\section{Introduction}
\label{sec:introduction}


High-purity germanium (HPGe) detectors are well suited to search for physics beyond the Standard Model 
such as neutrinoless double-beta ($0\nu\beta\beta$)
decay~\cite{LEGEND:2017cdu,0bnn:Review2019,Majorana:2019nbd,GERDA:2020xhi,LEGEND:2021bnm} 
and dark matter~\cite{CoGeNT:2013,SuperCDMS:2014,PhysRev:LimitWIMPS2018}.
In such rare-event searches, it is essential to suppress background events as much as possible.

The LEGEND experiment~\cite{LEGEND:2017cdu, LEGEND:2021bnm} will search for $0\nu\beta\beta$ decay in $^{76}$Ge 
using HPGe detectors based on the experience of the preceding 
experiments {\sc Gerda}~\cite{GERDA:2012qwd,GERDA:2020xhi} and {\sc Majorana Demonstrator}~\cite{Majorana:2013cem,Majorana:2019nbd}.
In these experiments, a substantial part of the background events with observed energies close to the $Q$-value of the $0\nu\beta\beta$ decay ($Q_{\beta\beta}=2039$\,keV)~\cite{0bnn:Review2019} originates from events near the surface of the detectors~\cite{GERDA:2019cav}.
Pulse-shape discrimination techniques have been developed to reject such surface events~\cite{GERDA:2022ixh}, including a dedicated method to identify events originating from the large passivated surfaces of the point-contact detectors of the {\sc Majorana Demonstrator}~\cite{Majorana:2020xvk}.
A substantial further reduction of background is essential for LEGEND to reach its goal of increasing the half-life sensitivity to $0\nu\beta\beta$ decay of $^{76}$Ge to $10^{28}$\,years and beyond~\cite{LEGEND:2021bnm}. 
A deep understanding of the HPGe detectors and their signal pulses is essential to achieve~this.

A study of alpha events on the passivated surface of the segmented true-coaxial n-type HPGe detector, "Super-Siegfried"~\cite{Abt:2016trw}, was performed in the GALATEA test facility~\cite{Abt:2014bpa} at the Max Planck Institute for Physics in Munich.
Such events are an important background to $0\nu\beta\beta$ decay searches when charge trapping reduces the observed energy to values within the region of interest around $Q_{\beta\beta}$~\cite{GERDA:2019cav}.

For the first time, the dependence of charge trapping on the orientation of the crystal axes is presented. 
First attempts to simulate surface effects with
the open-source software package \emph{SolidStateDetectors.jl}~\cite{Abt:2021mzq} are also shown.

In general, all contacts of germanium detectors are completely coated with a thin metal layer, i.e. are fully metalised.
However, the segments of the detector Super-Siegfried 
were originally not fully metalised but were only partially metalised on a dedicated contact area.
The results of studies~\cite{Abt:2016trw} where data were taken with partially metalised contacts 
were compared to results from data taken after the segments became fully metalised.
The influence of the extent of the metalisation is demonstrated.


\section{Experimental setup and data taking}
\label{ExpSetup}

\begin{figure*}[ht]
    \centering
    \includegraphics[width=1\textwidth]{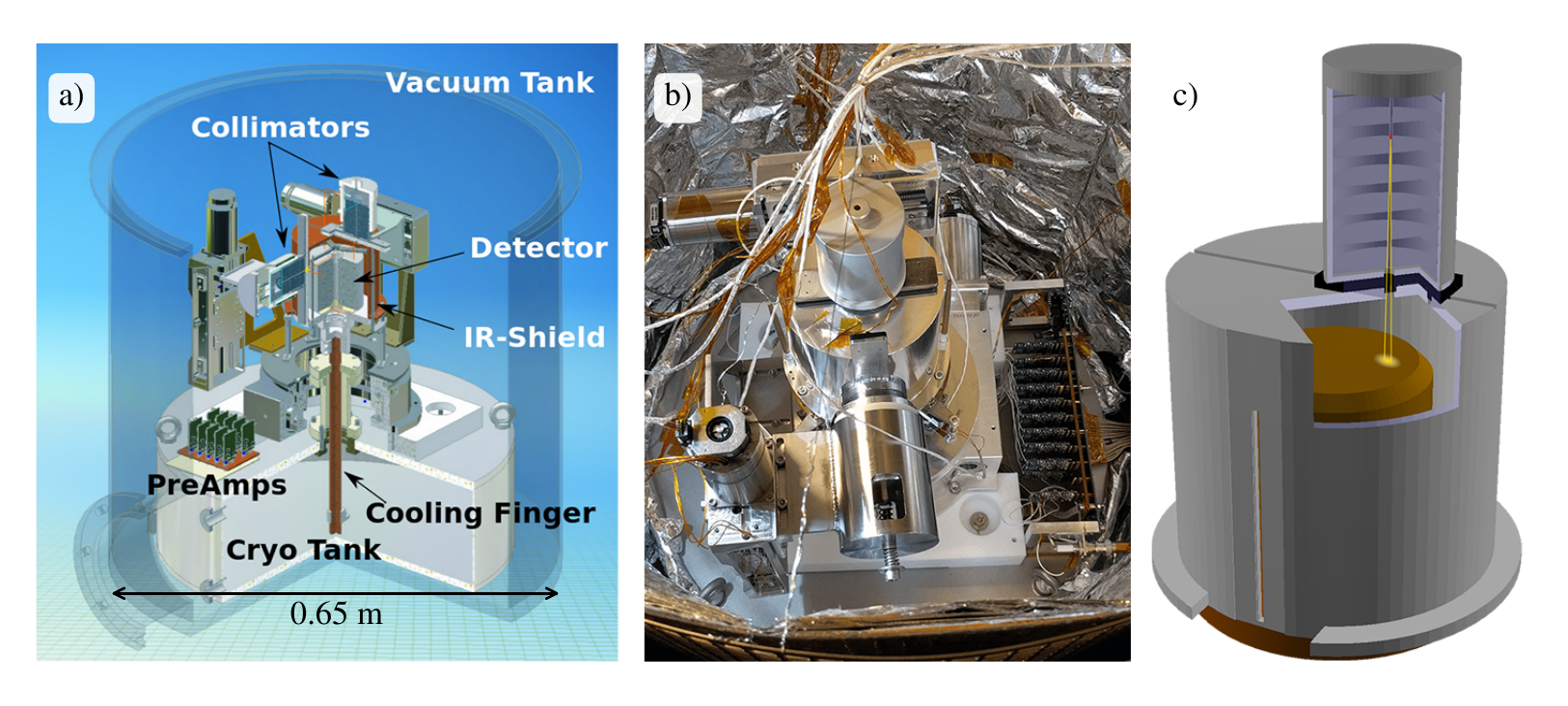}
    \caption{a) An illustration adapted from~\cite{Abt:2014bpa} and b) a photo of the vacuum chamber of GALATEA with the main components marked in a). c)~Schematic of the central part of GALATEA with the infrared (IR) shield and the top collimator and an illustration of the beam-spot.}
    \label{fig:galatea}
\end{figure*}

GALATEA is based on a vacuum chamber in which the top and side surfaces of a detector 
can be scanned using a system of three motorised stages, see Fig.~\ref{fig:galatea}.
The HPGe detector is mounted on a holder, which is cooled via a copper cooling finger 
submerged in an integrated liquid-nitrogen tank.
The fill level of the tank is monitored and the internal tank is 
automatically refilled through a connected dewar.

The special feature of GALATEA is that the path between the detector and the source is free of material, facilitating the use of alpha sources.
Two sources can be mounted in two tungsten collimators focusing the radiation onto the mantle and top plate of the detector, respectively.
The collimators are moved by linear stages and slide along slits on the side and top of a silver-coated copper hat serving as an infrared shield.
A rotational stage moves these components around the detector, such that almost any point on the detector top surface or mantle can be irradiated.
Significant improvements of the setup since previous studies~\cite{Abt:2016trw},
such as stable long-time operation, automatised data taking and the use of open alpha sources, allowed for more detailed studies of charge trapping.

\begin{figure}[hb]
    \centering
    \includegraphics[width=\columnwidth]{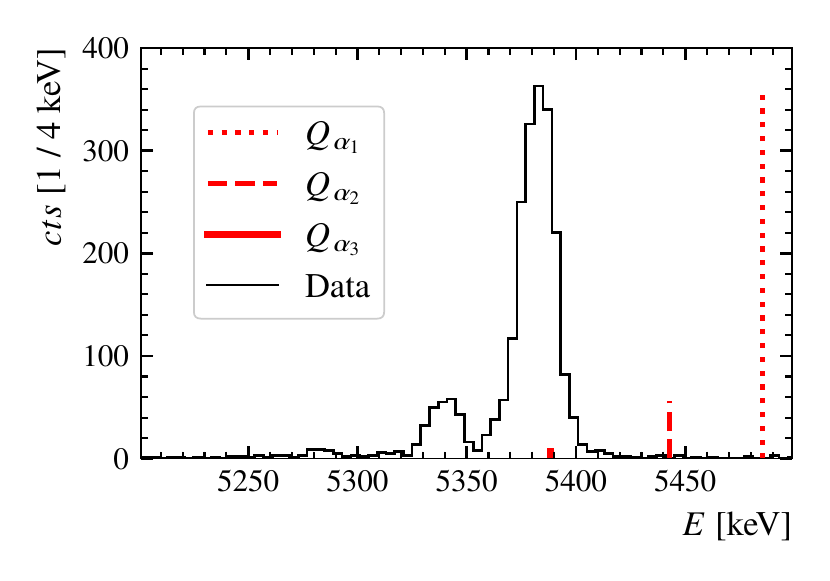}
    \caption{$Q$-values (vertical lines) and measured alpha energy spectrum (histogram) in response to $^{241}$Am irradiation onto the side of the Super-Siegfried detector for one hour. 
    The height of the vertical lines represent the branching ratio of the respective decay.}
    \label{fig:Am241_SideAlphaPeaks}
\end{figure}

Two open alpha $^{241}$Am sources, each with an activity of 74\,kBq, were installed in the two collimators allowing for simultaneous scans of the side and the top of the detector.
The isotope $^{241}$Am decays into $^{237}$Np with dominant alpha energies of $Q_{\alpha_1}=5485.6$\,keV, $Q_{\alpha_2}=5442.8$\,keV and $Q_{\alpha_3}=$\nolinebreak~5388.2\,keV~\cite{NuclearDataBase:1999}.
In addition, $^{237}$Np emits 59.5~keV gammas.
The alphas lose their energy within 20\,\textmu m of germanium with the Bragg peak at
17\,\textmu m~\cite{Hauertmann2021:PhDThesis} and create events directly underneath the surface of the detector.
Figure~\ref{fig:Am241_SideAlphaPeaks} shows the observed spectrum in response to irradiation from the side.
The reduction of the observed energies with respect to the emitted energies corresponds to the contact thickness on the side of the detector.
The width of the observed lines arises from self-absorption in the source.

Super-Siegfried is a cylindrical true-coaxial n-type segmented HPGe detector produced by Canberra France, now Mirion Technologies~\cite{PhD:Irlbeck2014}.
The detector has a height of 70\,mm, an outer radius of 37.5\,mm and a borehole with a radius of 5\,mm.
At the top and the bottom, the borehole widens to a radius of 10\,mm within about 3\,mm.
The impurity density as stated by the manufacturer of the crystal is $0.44\times10^{10}$\,cm$^{-3}$ at the top and $1.30\times10^{10}$\,cm$^{-3}$ at the bottom.
The recommended operational voltage is 3000\,V.

\begin{figure*}
    \centering
    \includegraphics[width=\textwidth]{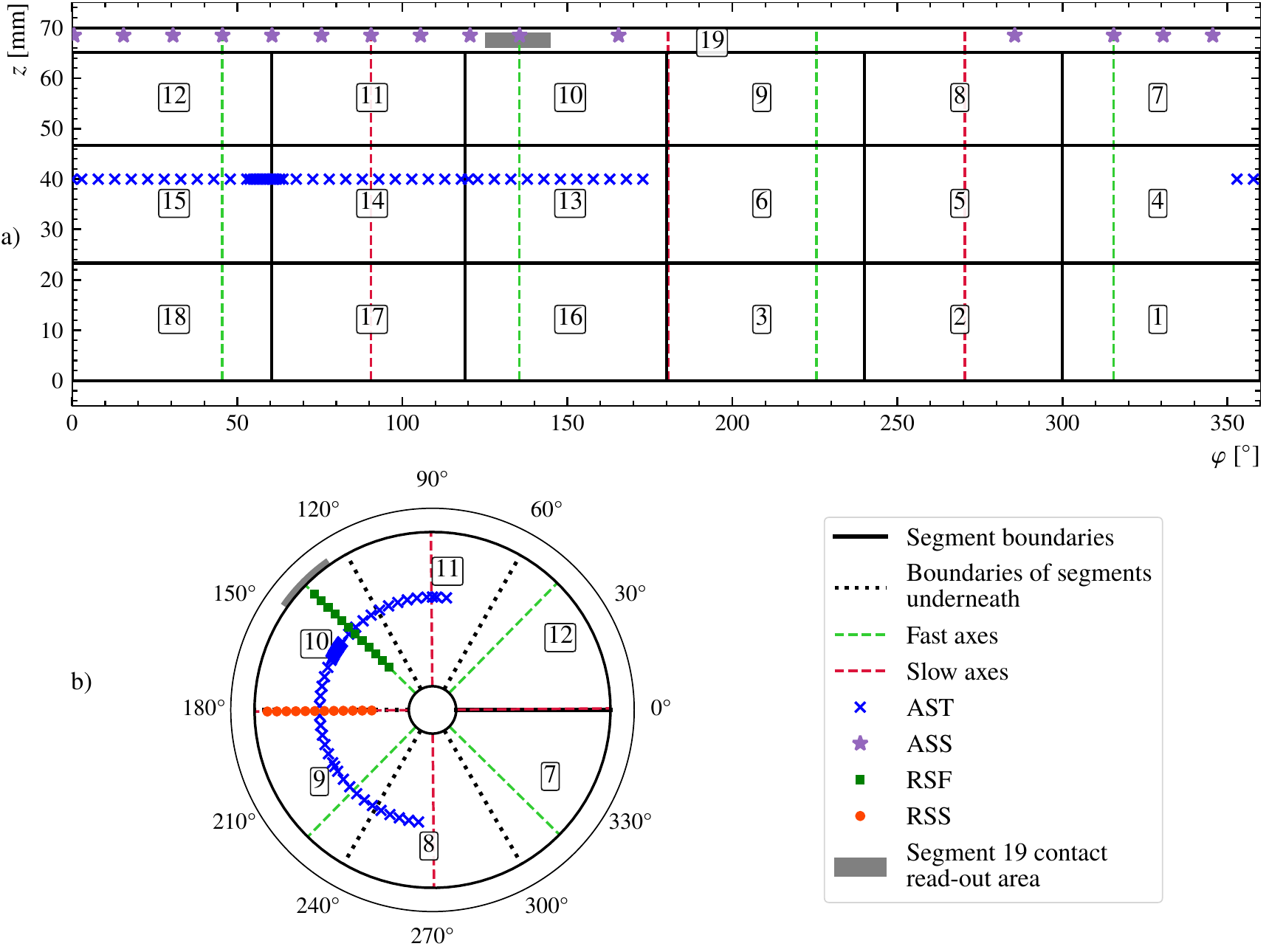}
    \caption{a) Side view: The segment structure of the mantle of Super-Siegfried with dimensions and reference coordinate system.
    b) Top view: The positions of the measurements on the top surface are marked.
    Also shown are the crystal axes in a) and b). 
    The framed numbers denote the segments numbers, in b) for the segments underneath segment~19.}
    \label{fig:scans}
\end{figure*}
\begin{figure}[ht]
    \centering
    \includegraphics[width=\columnwidth]{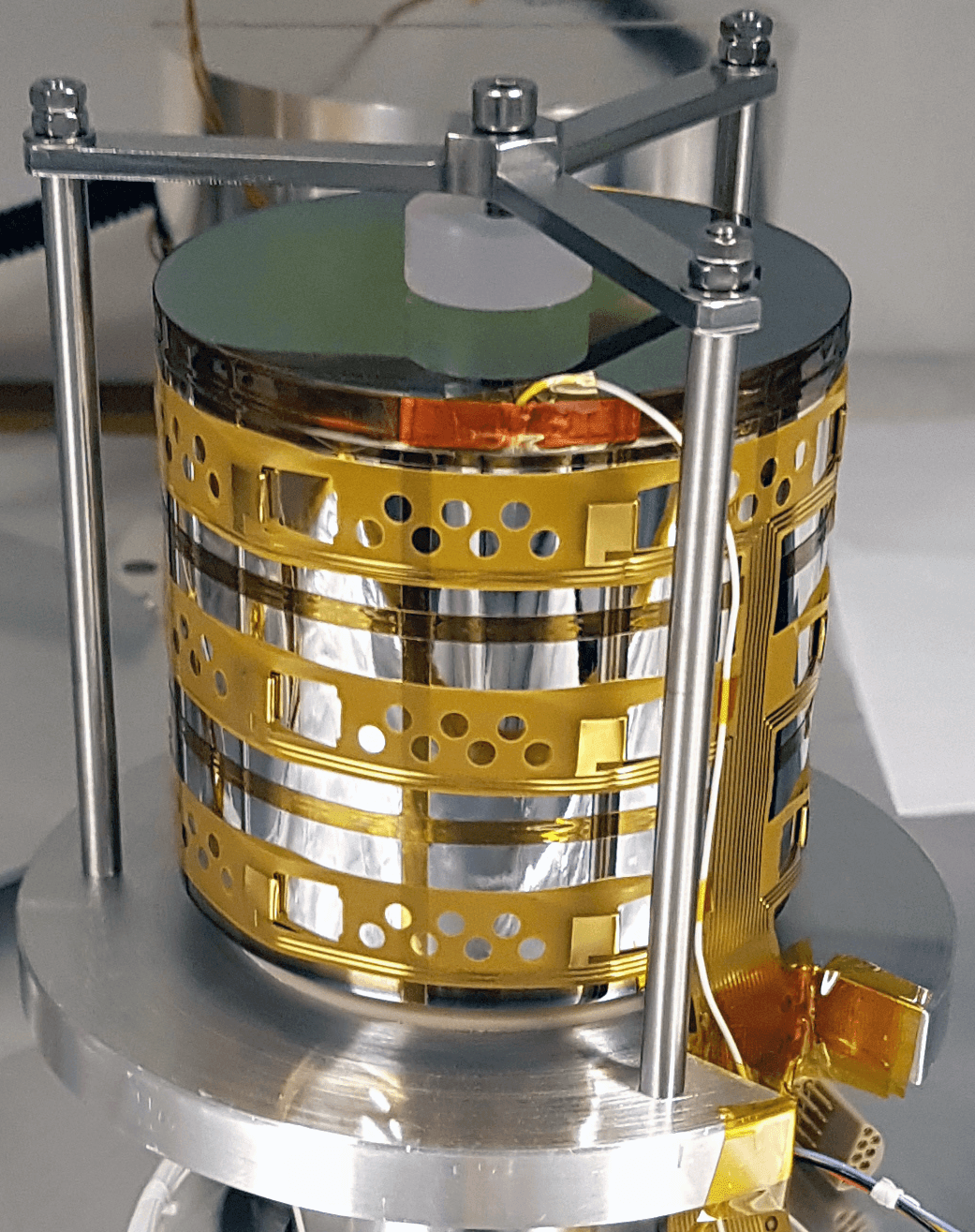}
    \caption{Picture of the germanium detector Super-Siegfried taken after all segments were fully metalised.
    The Kapton printed-circuit-board and the holding structure are also visible.}
    \label{fig:PicSuSie}
\end{figure}

The lithium-drifted inner surface of the borehole (not including the area of the widening) is the only n$^+$ contact, the so-called core contact.
The outer mantle is segmented into 19 p$^+$ contacts produced via boron implantation,
see Fig~\ref{fig:scans}a.
The 18 lower segments are ordered in a $6\times3$ geometry in $\varphi$ and $z$.
These segments are read-out at their centres via 
one Kapton printed-circuit-board~\cite{Abt:2007rf}.
An additional segment located above these 18 segments, segment~19, has a height of 5\,mm and is not segmented in~$\varphi$.
It is read out via a cable at $\varphi_{\mathrm{ro}}\approx135^{\circ}$, 
see Fig.~\ref{fig:scans}.
Prior to the full metalisation, the read-out cable for segment~19 was connected at about the same 
location.
The detector, as ready to be mounted on the cooling finger in GALATEA, is depicted in Fig.~\ref{fig:PicSuSie}.
The fairly thin segment~19 was designed to study events on the passivated surface of the top end-plate.
Charges drifting towards the segment~19 contact also create strong so-called mirror pulses in the segments underneath, which facilitate a detailed pulse-shape analysis.

The positions of the two sources were calibrated by moving the collimators and identifying the segment boundaries
and the crystal axes as well as the relative position of the holding structure~\cite{Hauertmann2021:PhDThesis}.
Figure~\ref{fig:scans}b shows the position of the $\langle100\rangle$ and $\langle110\rangle$ crystal axes, 
referred to as fast and slow axes, respectively, as the charge carriers drift faster and slower along these axes in germanium.
All events recorded for a given position of the sources are from here on called a measurement. 
All positions refer to the positions of the centre of the beam-spot on the detector surface.
The individual measurement positions on the top and side of the detector are also shown in Fig.~\ref{fig:scans}.
They comprise two radial scans, RSF along a fast axis and RSS along a slow axis, and 
two rotational (azimuthal) scans, AST from the top at a fixed radius of $23.8$\,mm and
from the side at $z=40$\,mm and ASS only from the side on segment~19.
The index $j$ is used throughout this paper to refer the $j$-th measurement of the respective scan.
Each measurement of these four scans lasted one hour.
In addition, a five hour long background measurement (BG) was performed where the sources did not irradiate the detector.

\begin{figure}[ht]
    \centering
    \includegraphics[width=\columnwidth]{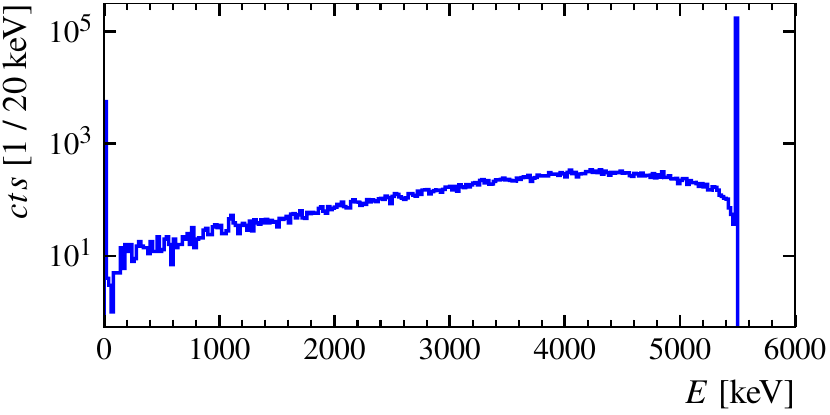}
    \caption{Energy spectrum of the alphas reaching the detector as predicted
    by {\sc Geant4} for monoenergetic 5.485\,MeV alphas from the source volume
    directed downwards with an opening angle of 3 degrees.}
    \label{fig:geant4_energy_spectrum}
\end{figure}

\begin{figure}[ht]
    \centering
    \includegraphics[width=\columnwidth]{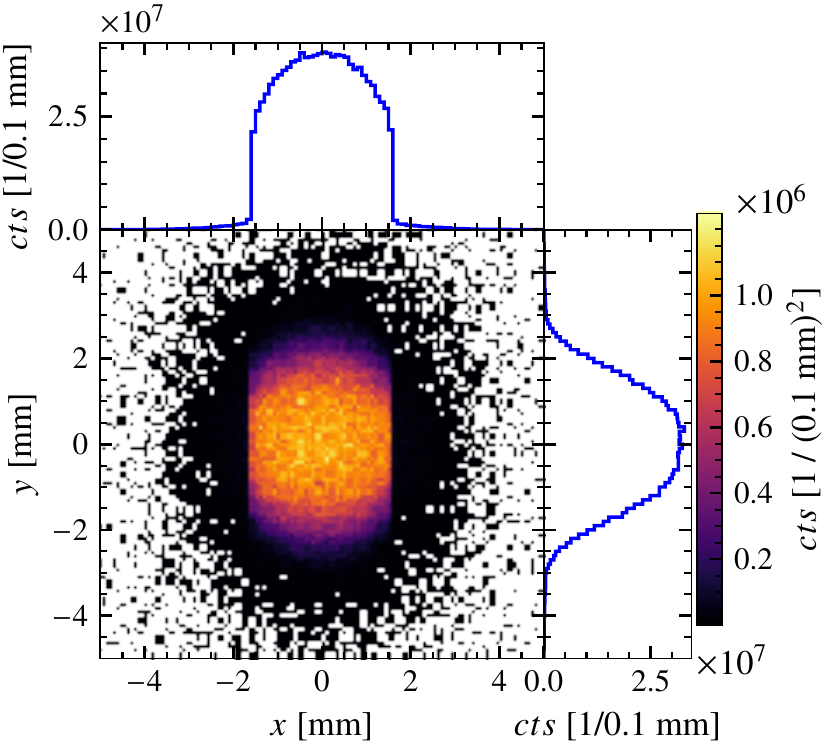}
    \caption{Distributions of energy-weighted energy depositions
        of the alphas reaching the detector as predicted
    by GEANT4 for monoenergetic 5.485\,MeV alphas from the source volume
    directed downwards with an opening angle of 3 degrees.
        The y-axis is parallel to the radial direction in the cylindrical coordinate system of the detector, 
        see Fig.~\ref{fig:scans}b.}
    \label{fig:geant4_top_beamspot}
\end{figure}

The alpha beam from the top as provided by the setup was simulated with
{\sc Geant4}~\cite{Geant4:2003} to investigate the incident spectrum and beam-spot on the top surface.
A simulation of monoenergetic 5.485\,MeV alphas from the source in the top collimator showed that the resulting spectrum has
a tail towards low energies, see Fig.~\ref{fig:geant4_energy_spectrum}. 
This tail is populated by alphas which loose energy
through interactions with the wall of the collimator or the edge of the
slit in the infrared shield.
The shape and population of the beam-spot is shown in Fig.~\ref{fig:geant4_top_beamspot}.
The sharp cut-off, perpendicular to the radial direction, is caused
by the additional collimation provided by the infrared shield.

For a single event, 20 pulses (from the core and all 19 segments), each with 5000 samples, 
were recorded with a sampling rate of 250\,MHz by two 16-channel Struck SIS3316~\cite{STRUCK} analog-to-digital converter units.
The index $i \in [0, 19]$ is used throughout this paper to specify the channel.
All devices of GALATEA were controlled and monitored via one software package, which allowed for automated scans of the detector.
The trigger was provided by the core channel connected to the first unit which forwarded its trigger signal to the second unit via a short cable.
The constant time delay of the second unit was taken into account in the offline pulse processing~\cite{Hauertmann2021:PhDThesis}.

\begin{figure}[htbp]
    \centering
    \includegraphics[width=\columnwidth]{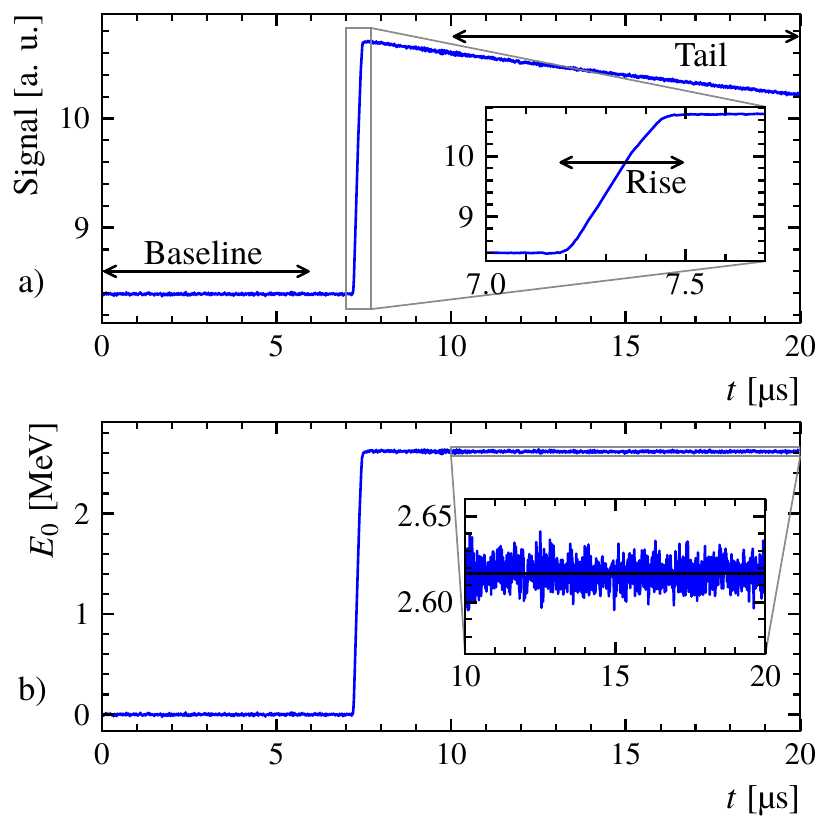}
    \caption{a) Typical raw pulse from the core channel of Super Siegfried.
    The three windows of the pulse are indicated by arrows and labelled.
    b) The same pulse after calibration and cross-talk correction.
    The horizontal line in the inset in b) indicates the mean of the tail of the pulse.}
    \label{fig:core_pulse}
\end{figure}

A typical raw (prior to any processing) pulse, from the core channel is shown in Fig.~\ref{fig:core_pulse}a.
This included a correction for the decay of charge in the preamplifier of each channel $i$~\cite{Hauertmann2017:MasterThesis}.
After baseline subtraction, an exponential function, $e^{-t/\tau_i}$,
was fitted to the part of each pulse following the rise, the so-called tail.
For all measurements, the characteristic decay constants of each preamplifier, 
$\tau^{c}_{i}$, were determined by fitting a Cauchy function to the $\tau_i$-distributions for bulk events~\cite{Hauertmann2017:MasterThesis}.
The pulse shown in Fig.~\ref{fig:core_pulse}a is shown in
Fig.~\ref{fig:core_pulse}b after cross-talk correction and calibration.
All pulses were corrected for linear cross-talk and calibrated for each individual measurement with an automated procedure~\cite{Hauertmann2017:MasterThesis}.

\section{Alpha event characteristics and selection}
\label{sec:selection}

\begin{figure}[ht]
    \centering
    \includegraphics[width=\columnwidth]{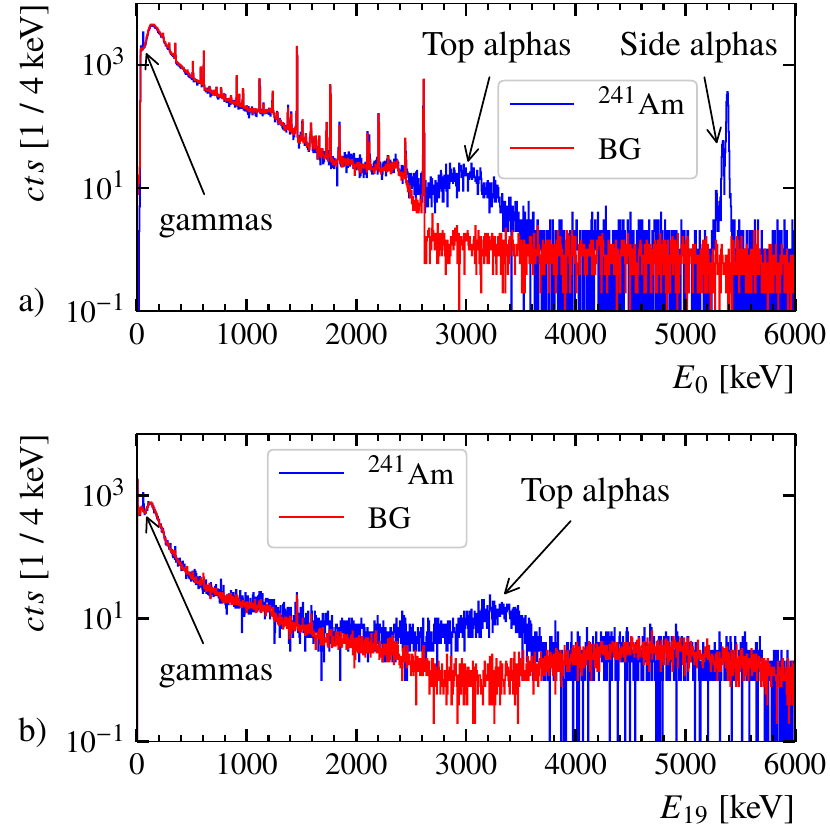}
    \caption{a) Core, $E_{0}$, and b) segment~19, $E_{19}$, spectra of the background measurement BG and a typical $^{241}$Am measurement from AST with the top beam-spot at ($\varphi=257.9^{\circ}$, $r=23.8$\,mm) and the side beam-spot at ($\varphi=167.9^{\circ}$, $z=40.0$\,mm). The spectrum of the background measurement is normalised in time to the $^{241}$Am measurement.}
    \label{fig:CoreEnergySpectra_BG_and_Am241}
\end{figure}

The spectra of the calibrated core and segment~19 energies, $E_{0}$ and $E_{19}$, 
are shown in Fig.~\ref{fig:CoreEnergySpectra_BG_and_Am241} for a measurement from AST with both $^{241}$Am sources 
irradiating the detector. 
The signal of the alphas hitting the top and the side of the detector are well separated due to the different structure of the surface layers.
The respective backgrounds from natural radioactivity are also shown for comparison.

Alphas deposit their energy in a small volume ($<1$\,mm$^{3}$) and create so-called single-site events. 
Thus, the collecting segment for alpha events from the top is always segment~19 and
$E_0$ and $E_{19}$ should be equal to one of the $Q_{\alpha_{k}}$,~$k\in\nolinebreak\{1,2,3\}$.
However, Fig.~\ref{fig:CoreEnergySpectra_BG_and_Am241} shows that the observed energies, $E_{0}$ and $E_{19}$, 
of these top alpha events are massively reduced compared to $Q_{\alpha_k}$ and 
that these observed alpha energies form wide distributions instead of three peaks. In addition, the shape and central position of the 
distributions are slightly different in the core and segment~19.
The reduction of energy, $\mathcal{O}$(2\,MeV), is partially caused by a so-called dead layer on top of the detector, 
where the created electron-hole pairs recombine immediately and, thus, do not contribute to the signal pulses.
However, a dead layer alone can neither explain the widths of the distributions nor the non-zero 
difference between $E_{0}$ and $E_{19}$.
Both observations can only be explained by the trapping of charge carriers during their drift through the detector. 

In this case, electrons or holes are trapped in the crystal and
do not reach a contact. After the other charge carriers are collected,
the trapped charge carriers still induce signals in the contacts.
This results in reduced and unequal values of $E_0$ and $E_{19}$,
depending on the position of the trapping.
In non-collecting segments, so-called truncated mirror pulses~\cite{Abt:2016trw}
are observed where the signal does not return to the baseline.

\begin{figure}[ht]
    \centering
    \includegraphics[width=\columnwidth]{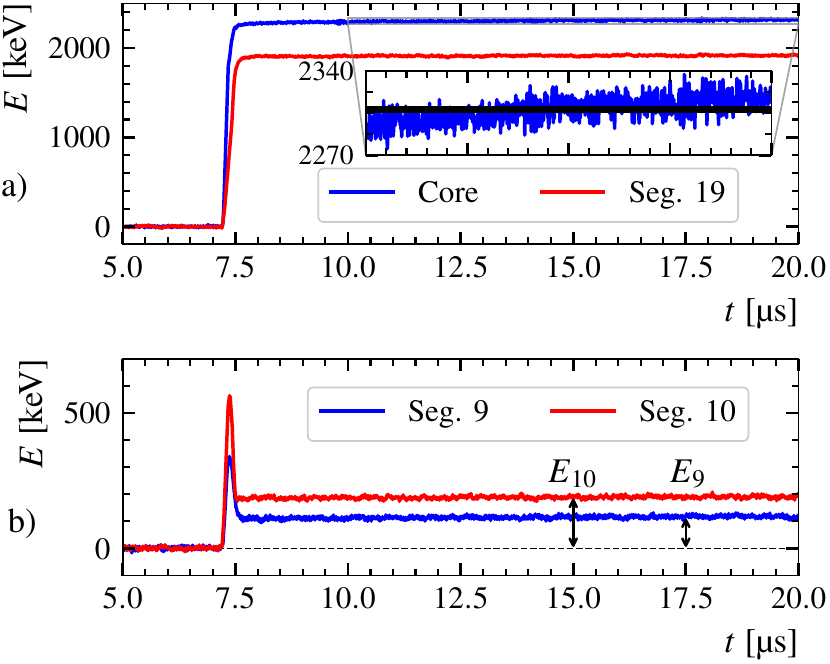}
    \caption{Pulses of a typical event with net hole trapping of the a) core and segment~19 and b) the two closest segments 9 and 10 
    of a measurement from RSS at $\varphi=180.6^{\circ}$ and $r=18.8$\,mm. 
    The observed energies, $E_{9}$ and $E_{10}$, are also indicated. 
    The horizontal line in the inset in a) is at the mean of the tail of the core pulse.}
    \label{fig:holetrapping}
\end{figure}

\begin{figure}[ht]
    \centering
    \includegraphics[width=\columnwidth]{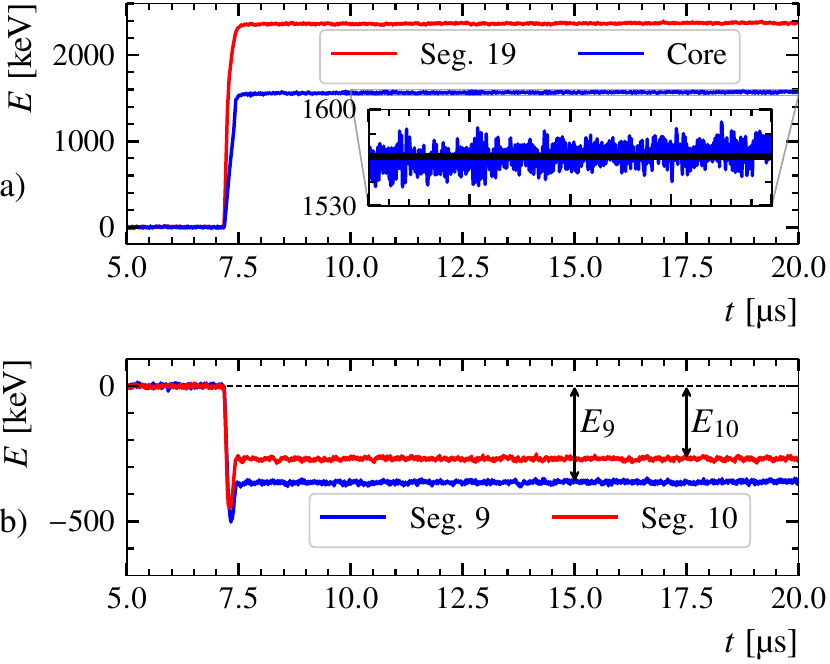}
    \caption{Pulses of a typical event with net electron trapping of the a) core and segment~19 and b) the two closest segments~9 
    and 10 of the measurement from RSS at $\varphi=180.6^{\circ}$ and $r=32.8$\,mm. 
    The observed energies, $E_{9}$ and $E_{10}$, are also indicated.
    The horizontal line in the inset in a) is at the mean of the tail of the core pulse.}
    \label{fig:electrontrapping}
\end{figure}

The characteristic pulses from an event with net
hole trapping, $E_{19} < E_{0}$, are shown in Fig.~\ref{fig:holetrapping}.
The collecting segment~19 shows a lower energy than the core and the two segments 
underneath, segments 9 and 10, show positive truncated mirror pulses indicating net hole trapping.
The characteristic pulses from an event with net electron trapping, $E_{0} < E_{19}$, are shown in Fig.~\ref{fig:electrontrapping}, where negative truncated mirror pulses are observed in segments 9 and 10.

Due to the feature of truncated mirror pulses, alpha events affected by charge 
trapping can also appear as so-called multi-segment events even though they are single-site events.
Thus, a strict single-segment~19 cut, $E_{0}\approx E_{19}$, cannot be used to filter out 
events induced by environmental gammas, which are often multi-segment events as the gammas 
predominantly perform Compton scattering at these energies.
However, in contrast to gamma induced events, the energies measured in the other segments should never be 
larger than $E_{19}$ for events induced by alphas from the top. 
Therefore, a softer single-segment~19 cut was defined to filter out gamma events:

\begin{itemize}
    \item[$\bullet$] S-cut: $E_{19} > \sum_{i=1}^{18}E_{i}$\,.
\end{itemize}

Alpha events affected by charge trapping feature another characteristic:
After the correction for the specific decay of charge in the preamplifiers, pulses from "normal" events have a flat tail, see Fig.~\ref{fig:core_pulse}b.
However, it has been shown that the pulses of alpha events on passivated surfaces have positive tail-slopes~\cite{Majorana:2020xvk,Hauertmann2021:PhDThesis}
after the preamplifier specific correction using the decay constants $\tau^{c}_{i}$.
These positive tail-slopes are visible in Fig.~\ref{fig:holetrapping}a and Fig.~\ref{fig:electrontrapping}a.
The possible physical origins of the non-zero tail slopes of alpha-event pulses are a slow release of trapped charge carriers or a very slow drift of a part of the charge cloud due to a very low mobility near the surface.
A positive tail-slope corresponds to a seemingly increased decay constant of the pulse, $\tau_{i}$, in comparison to $\tau^{c}_{i}$.
Thus, alpha events can be identified prior to any pulse processing or cross-talk correction by comparing 
$\tau_i$ with $\tau^{c}_{i}$.

\begin{figure}[!b]
    \centering
    \includegraphics[width=\columnwidth]{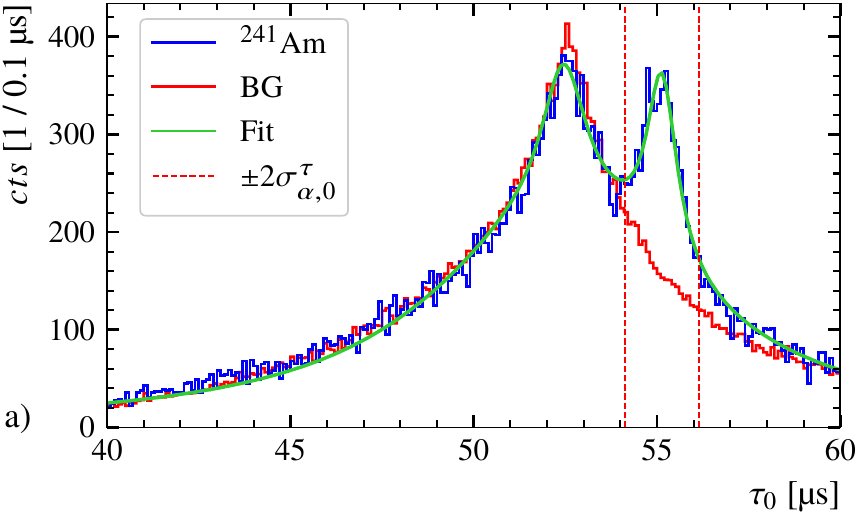}
    \includegraphics[width=\columnwidth]{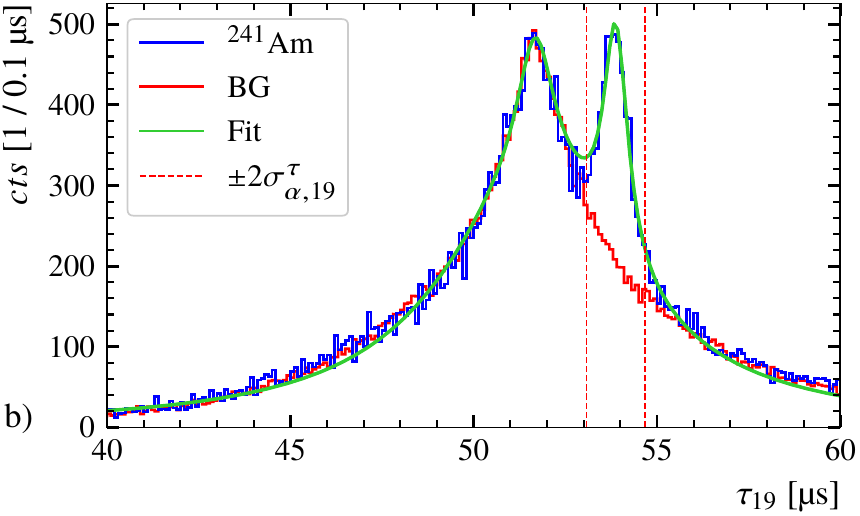}
    \caption{Distributions of a) $\tau_0$ and b) $\tau_{19}$ for events selected with the S-cut from AST at $\varphi=257.9^{\circ}$ and $r=23.8$\,mm and from BG.
    The solid line is the function $\mathcal{T}$, see Eq.~\eqref{eq:TauModelFunction}, fitted to the distributions of the $^{241}$Am measurement. 
    The $\pm2\sigma$ interval, used for the $\tau$-cut, is indicated by the vertical dashed lines.}
    \label{fig:TDCDists_and_fits}
\end{figure}

Typical $\tau_i$ distributions for events passing the S-cut are shown in Fig.~\ref{fig:TDCDists_and_fits} for a measurement from AST and BG.
The $\tau_i$ distributions can be modelled by a superposition of Cauchy 
functions ($\mathcal{C}$) via the function $\mathcal{T}(\mathcal{P}_{\mathcal{T}})$ with the parameters~$\mathcal{P}_{\mathcal{T}}$
\begin{linenomath}
\begin{align}
        \mathcal{P}_{\mathcal{T}} =&~\{ \mu^{\tau},\sigma^{\tau}_1,\sigma^{\tau}_2,A^{\tau}_1,A^{\tau}_2,
            A^{\tau}_{\alpha},\mu^{\tau}_{\alpha},\sigma^{\tau}_{\alpha} \} \, , \\
        \mathcal{T}(\mathcal{P}_{\mathcal{T}}) =&~
            A^{\tau}_1 \cdot \mathcal{C}(\mu^{\tau}, \sigma^{\tau}_1) + 
            A^{\tau}_2 \cdot \mathcal{C}(\mu^{\tau}, \sigma^{\tau}_2)\,+  \label{eq:TauModelFunction} \\
            &~A^{\tau}_{\alpha} \cdot \mathcal{C}(\mu^{\tau}_{\alpha}, \sigma^{\tau}_{\alpha}) \notag \, ,
\end{align}
\end{linenomath}
where $\mu^{\tau},\sigma^{\tau}_1,\sigma^{\tau}_2,A^{\tau}_1$ and $A^{\tau}_2$ 
model the underlying background from "normal" events and 
$A^{\tau}_{\alpha},\mu^{\tau}_{\alpha}$ and $\sigma^{\tau}_{\alpha}$ model the
second peak which is associated with alpha events.

In order to study charge trapping in detail, the events induced by alphas from the top source are 
selected with the S-cut and a second cut:
\begin{itemize}
\item[$\bullet$] $\tau$-cut: $\tau_0 \in [\mu^{\tau}_{\alpha,0} \pm 2\sigma^{\tau}_{\alpha,0}$] and 
                 $\tau_{19} \in [\mu^{\tau}_{\alpha,19} \pm 2\sigma^{\tau}_{\alpha,19}$]\, .
\end{itemize}


\begin{figure}[htbp]
    \centering
    \includegraphics[width=\columnwidth]{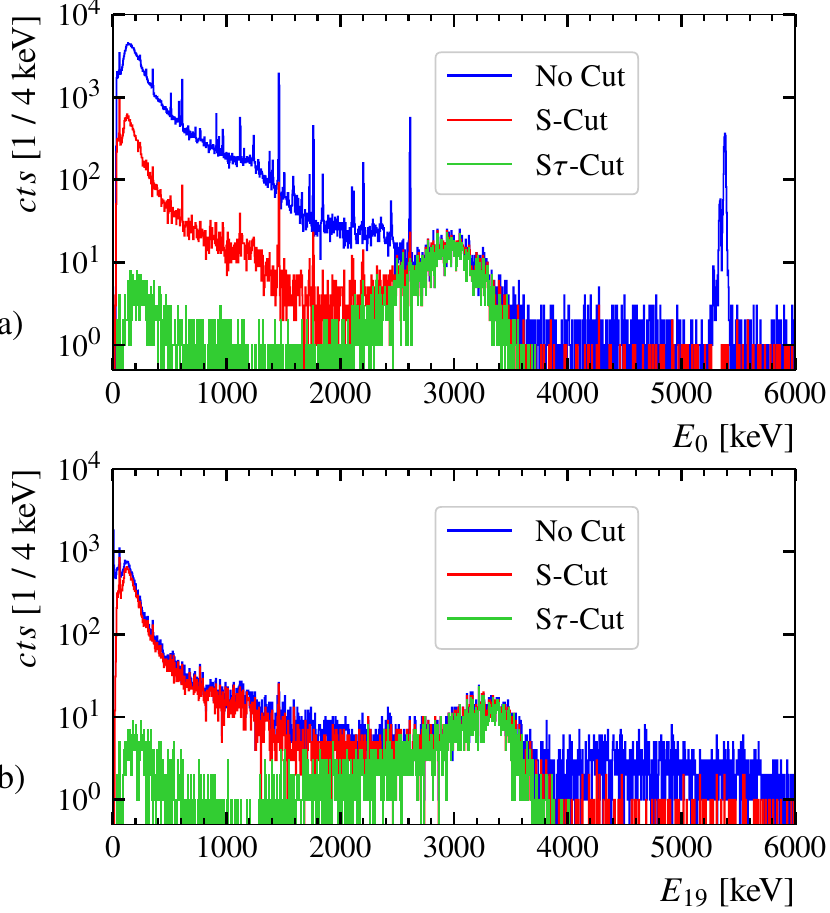}
    \caption{Spectra of a) $E_{0}$ and b) $E_{19}$ before and after the application of the S-cut and S$\tau$-cut for a measurement of the scan AST with the top beam-spot at ($\varphi=257.9^{\circ}$, $r=23.8$\,mm) and the side beam-spot at ($\varphi=167.9^{\circ}$, $z=40.0$\,mm).}
    \label{fig:CoreEnergySpectra_TopAlphaCuts}
\end{figure}

Figure~\ref{fig:CoreEnergySpectra_TopAlphaCuts} shows the $E_{0}$ and $E_{19}$ spectra before the alpha selection cuts, after applying the S-cut and after applying the additional $\tau$-cut (S$\tau$-cut).
The S-cut removes events with alphas entering the side of the detector and suppresses 
gamma events, for which most of the energy is deposited outside of segment~19.
After the S$\tau$-cut, only alpha events on the top remain.
Above 1000\,keV, they form a broad peak.
It has been cross-checked that the events in the energy region of the broad alpha peak indeed form the secondary peak observed in the $\tau$ distribution, confirming that the $\tau$-cut selects alpha events~\cite{Hauertmann2021:PhDThesis}.
Similarly, analysing the mirror pulses from the neighbouring segments also shows that the S$\tau$-cut is effective in selecting alpha events with charge trapping.
These checks verify that the core pulse alone is sufficient to select alpha events.
Thus, the rejection of events based only on positive tail slopes, is confirmed to be an effective way to reduce the background for $0\nu\beta\beta$ decay searches.
It should be noted that signal events very close to the surface might also be rejected.

\section{Location dependence of charge trapping}

\begin{figure}[ht]
    \centering
    \includegraphics[width=\columnwidth]{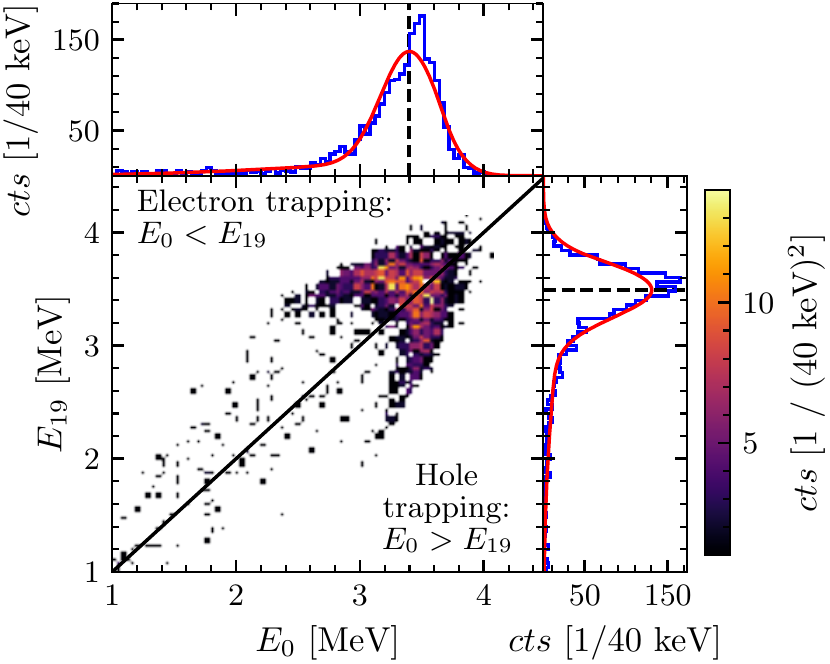}
    \caption{Correlation between core and segment~19 energies of alpha events selected as described in 
    Sec.~\ref{sec:selection} from a measurement of the scan AST at $\varphi=232.9^{\circ}$ and $r=23.8$\,mm. 
    Also shown are the two marginalisations.
    The dashed (black) lines mark the position of the parameter $E^{0/19}_{\mu,j,\mathrm{obs}}$ for this 
    measurement $j$ of the respective fitted function $\mathcal{M}$, see Eq.~\eqref{eq:AlphaEnergyBumpModelFunction}, shown as a solid (red) line.}
    \label{fig:trapping2d}
\end{figure}

The correlation between $E_{0}$ and $E_{19}$ is shown in Fig.~\ref{fig:trapping2d}
for the selected alpha events from the measurement at $\varphi=\nolinebreak257.9^{\circ}$ and $r=23.8$\,mm from AST.
Events at this radial position are subject to more electron trapping and higher energies are observed in segment~19 than in the core.
The $E_{0}$ and $E_{19}$ distributions are fitted via maximum likelihood estimation to find the most likely $E_{0}$ and $E_{19}$.
The function~$\mathcal{M}(E|\mathcal{P}_{\mathcal{M}})$, with the parameters $\mathcal{P}_{\mathcal{M}}$, 
defined as
\begin{linenomath}
\begin{align} 
    \mathcal{P}_{\mathcal{M}} =&~\{A,R,E_{\mu},\sigma_{1},\Delta E_{\mu},\sigma_{2},s\} \, ,\\
        \mathcal{M}(E|\mathcal{P}_{\mathcal{M}}) =&~A\cdot [(1 - R) \cdot \mathcal{N}(E|E_{\mu}, \sigma_{1}) \notag  \\ 
        & \,\,\,\,\,\,\,\,\,\,\, + R \cdot \mathcal{N}(E|E_\mu - \Delta E_{\mu}, \sigma_{2}) \label{eq:AlphaEnergyBumpModelFunction}\\ 
        & \,\,\,\,\,\,\,\,\,\,\, \cdot (\textrm{erf}(-s \cdot (E - (E_{\mu} - \Delta E_{\mu}))) + 1) ]  \notag 
\end{align}
\end{linenomath}
was chosen to model the distribution of observed energies of the selected events.
It consists of two Normal distributions~$\mathcal{N}$: the first, $\mathcal{N}(E_{\mu}, \sigma_{1})$, modelling the main peak and the second, $\mathcal{N}(E_\mu - \Delta E_{\mu}, \sigma_{2})$ modulated by an error function, modelling the low-energy tail of the distribution.
The parameter $E_{\mu}$ corresponds to the most likely observed energy of the alphas.

\begin{figure*}[htpb]
    \centering
    \includegraphics[width=\textwidth]{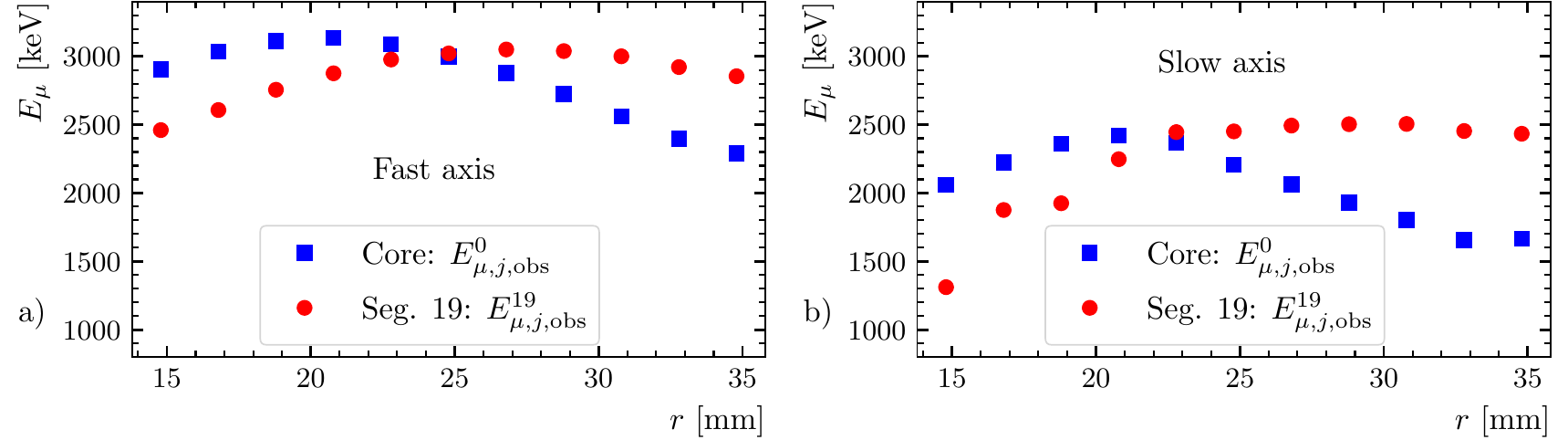}
    \caption{The fitted $E_{\mu,j,\mathrm{obs}}^{0}$ and $E_{\mu,j,\mathrm{obs}}^{19}$ of the measurements from a) RSF and b) RSS.
    Error bars indicating uncertainties would be hidden by the markers.}
    \label{fig:rdep}
\end{figure*}

The obtained values for the core, $E_{\mu,j,\mathrm{obs}}^{0}$ and segment~19, $E_{\mu,j,\mathrm{obs}}^{19}$, 
for the different measurements in the radial scans RSF (fast axis) and RSS (slow axis) are shown in Fig.~\ref{fig:rdep} as a function of $r$.
The uncertainties on the fitted values of $E_{\mu}$ are only a few keV.
These results confirm the trends observed previously~\cite{Abt:2016trw} that for events at smaller radii 
hole trapping dominates and moving towards segment~19 electron trapping becomes dominant.
This can be understood as a higher probability of charge trapping when the charge carriers have a longer drift path towards the contacts.
The result of the two scans reveal a dependence of the amount of charge trapping on the crystal axes orientation.
The energies observed in the scan along the fast axis are up to 1\,MeV higher than those in the scan along the slow axis.

\begin{figure*}[htpb]
    \centering
    \includegraphics[width=\textwidth]{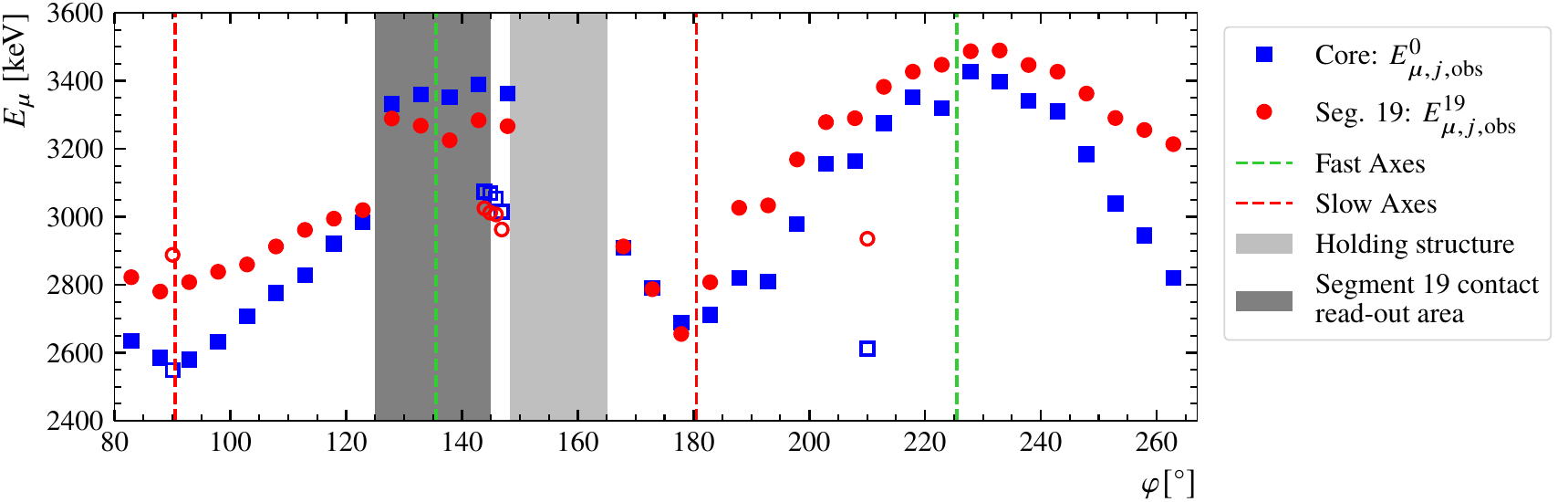}
    \caption{The fitted $E_{\mu,j,\mathrm{obs}}^{0}$ and $E_{\mu,j,\mathrm{obs}}^{19}$ of the measurements from AST at $r=23.8$\,mm. 
    Error bars indicating uncertainties would be hidden by the markers.
    Data points taken four months later are shown with open markers.
    }
    \label{fig:phidep}
\end{figure*}

The fitted values of $E_{\mu,j,\mathrm{obs}}^{0}$ and $E_{\mu,j,\mathrm{obs}}^{19}$ over $\varphi$ are shown in Fig.~\ref{fig:phidep}
for the measurements of the rotational scan AST at $r=23.8$\,mm.
There is a clear effect of the crystal axes on $E_{\mu,j,\mathrm{obs}}^{0}$ and $E_{\mu,j,\mathrm{obs}}^{19}$.
Near a fast axis, more energy is recorded, which would imply a thinner dead layer or less net trapping of charge carriers.
The latter is reasonable as the mobility for drifts along the fast axis is higher.
An effect of the crystal axes on the observed alpha energies is seen for the first time in GALATEA.
This was possible due to the increased amount of data compared to previous studies~\cite{Abt:2016trw}.

The open markers in Fig.~\ref{fig:phidep} show data points taken four months after the original scan. 
Several warming-cooling cycles probably changed the surface conditions influencing the amount of charge trapping.


\section{Simulation of charge trapping}

A dead layer alone can only equally reduce the alpha energy observed in the core and segment~19.
It can neither explain the observed differences between $E_{\mu,j,\mathrm{obs}}^{0}$ and $E_{\mu,j,\mathrm{obs}}^{19}$
nor the observed truncated mirror pulses in non-collecting segments.
The observed alpha energies can only be explained by taking charge trapping effects into account.
There is no established microscopic model available which includes charge trapping.
Here, a model consisting of three different parts is introduced to describe the radial 
dependence of $E_{\mu,j,\mathrm{obs}}^{0}$ and $E_{\mu,j,\mathrm{obs}}^{19}$.
The model was implemented using \emph{SolidStateDetectors.jl} (SSD)~\cite{Abt:2021mzq}, which is an open-source package to simulate solid-state detectors such as germanium detectors. 
It is written in the Julia programming language and it comprises the calculation of the electric field, $\mathbfcal{E}$, 
the weighting potentials~\cite{Shockley,Abt:2021mzq} of the contacts, $W_i$, 
the drift of the charge carriers through the semiconductor and the induced signals on the contacts.
The drift of the charge carriers is calculated in time steps by deriving a drift vector, $\mathbf{v}_d$, from the mobility, $\mu^{e/h}$, and 
the electric field at the current position of the respective charge carrier: $\mathbf{v}^{e/h}_d = \mu^{e/h} \cdot \mathbfcal{E}$. 
In each time step, the signal in each channel is determined by evaluating the 
weighting potential of all charge carriers at their current position.

The three parts of the model describing the physics underneath the passivated surface are: 
\begin{enumerate}
    \item a so-called dead layer,
    \item a so-called surface channel, 
    \item probabilistic charge trapping.
\end{enumerate}

The dead layer varying with $r$, $DL(r)$, is implemented for a thickness modelled as a quadratic function
with three parameters, $DL_1$, $DL_2$ and $DL_3$, which are the dead-layer thicknesses at the radii 12.8\,mm, 23.0\,mm and 34.8\,mm.
The allowed parameter space for these three parameters was chosen as [0,\,20]\,\textmu m which is motivated by the location of the Bragg peak for alphas as described in Sec.~\ref{ExpSetup}.
For all $r$, $DL(r)$ is forced to be $\geq 0$\,\textmu m.
The initial energy, $Q_{\alpha}^*(r)$, of the events is the reduced energy based on the 
thickness of the dead layer at the respective radial position.
The reduction is calculated by integrating the energy loss of alphas in germanium over the dead-layer thickness~\cite{Hauertmann2021:PhDThesis}.

The surface channel is implemented in the simulation as a virtual volume, in which $\mathbf{v}^{e/h}_d$ is 
modulated to ensure a drift parallel to the passivated top surface, 
\begin{linenomath}
\begin{equation}
    \mathbf{v}^{e/h}_{d,\mathrm{m}} = (\mathbf{v}^{e/h}_d \cdot \mathbf{e}_r) \cdot \mathbf{e}_r + 
        (\mathbf{v}^{e/h}_d \cdot \mathbf{e}_{\varphi}) \cdot \mathbf{e}_{\varphi},    
\end{equation}
\end{linenomath}
where $\mathbf{e}_r$ and $\mathbf{e}_{\varphi}$ are the radial and azimuthal unit vectors.
This virtual volume is a tube directly underneath the top surface with a height of 100\,\textmu m, ranging from $r = 10.05$\,mm to $r = 37.45$\,mm.
The height of the virtual volume is motivated by the fact that charge trapping was not observed for 59.5\,keV gammas also emitted by the $^{241}$Am source~\cite{Hauertmann2021:PhDThesis}. 
Most of these gammas interact with the germanium within about 1\,mm.
Thus, only events very close, $\mathcal{O}$(20\,\textmu m), to the surface seem to be affected by surface effects.
The modulation of the drift vector, which corresponds to a modulation of the mobility tensor, is justified as the 
assumption of an infinite crystal made in the derivation of the mobility tensors for electrons and holes is not valid this close to the surface.

\begin{figure}[htbp]
    \centering
    \includegraphics[width=\columnwidth]{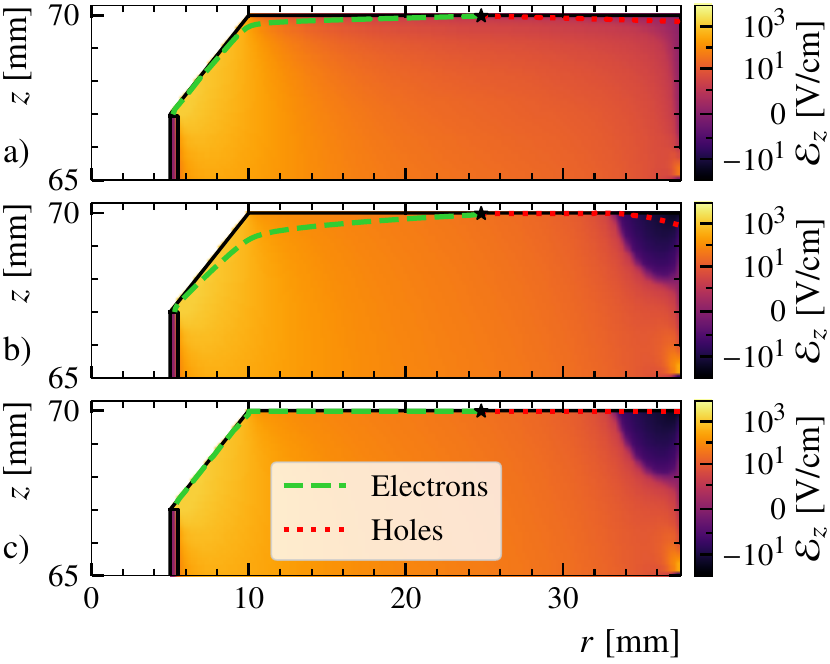}
    \caption{The $z$-component of the electric field, $\mathcal{E}_z$, in the top part of Super-Siegfried. 
    The dashed (dotted) lines are the drift trajectories of the electrons (holes) of an event simulated 
    at $r = 24.8$\,mm and $\varphi=135.6^\circ$ corresponding to one measurement position of the scan RSF 
    marked by the star.
    In a), the field simulation is limited to the volume of the detector. 
    In b) and c), the field simulation includes the infrared shield and detector base plate of GALATEA. 
    In c), the drift model is modified, forcing the charge carriers to drift along the surface.}
    \label{fig:fieldpath}
\end{figure}

It is also clear that the environment of the detector has to be taken into account in the simulation 
when handling the drift of charge carriers close to surfaces.
If the field simulation is limited to the volume of the detector, reflecting boundary conditions 
are typically used at surfaces like the passivated top plate. 
This, however, forces the $z$-component of the electric field, $\mathcal{E}_z$, to be zero at the surface, which is shown in Fig.~\ref{fig:fieldpath}a.
Even though this actually leads to the desired drift along the surface, it would incorrectly shift the cause for it from the mobility tensor to the electric field.
Thus, in order to learn something about the mobility tensor close to surfaces, one has to use the correct electric field, which can have a non-zero $z$-component at these surfaces as shown in Fig.~\ref{fig:fieldpath}b in which 
also the unmodulated drift paths for one event at $r = 24.8$\,mm just underneath the top surface are shown.
The modulated drift paths of the event are shown in Fig.~\ref{fig:fieldpath}c.

The probabilistic trapping of charge carriers is introduced in each step of the drift inside the virtual volume underneath the top plate as a probability for the charge carrier (electron or hole) to get trapped, $p_{t}^{e/h}$.
It is calculated based on the step duration, $\Delta t = 1$\,ns, on the current modulated drift velocity, $|\mathbf{v}^{e/h}_{d,\mathrm{m}}|$, 
and on the probability parameters $p_{t,0}^e$ and $p_{t,0}^h$ for electrons and holes:
\begin{linenomath}
\begin{equation}
    p_{t}^{e/h}(\Delta t, |\mathbf{v}^{e/h}_{d,\mathrm{m}}|) = 1 - \exp(-p_{t,0}^{e/h} \cdot \Delta t / |\mathbf{v}^{e/h}_{d,\mathrm{m}}|)\hspace{10pt}.
\end{equation}
\end{linenomath}
The allowed parameter space for $p_{t,0}^e$ and $p_{t,0}^h$ is [0,\,$\infty$]\,\textmu m/ns$^2$ ensuring $p_{t}^{e/h}(\Delta t, |v_d|) \in [0, 1]$. In each step, a random number between 0 and 1 is generated. If the generated number is below the $p_{t}^{e/h}$ of that step, the charge carrier is 
permanently stopped (trapped) at its current position.

A Bayesian fit was performed with the software \emph{BAT.jl}~\cite{Schulz:2021BAT} to the data points of RSF to determine the best values for the five parameters $DL_1$, $DL_2$, $DL_3$, $p_{t,0}^e$ and $p_{t,0}^h$. 
Flat priors were assumed for all five parameters.

For the fit, one event per set of parameters was simulated for each location.
Each event was simulated with 2000 electron-hole pairs, $N_{\mathrm{ep}} = 2000$, and the signal induced by each charge carrier was weighted with $Q_{\alpha}^*(r)/N_{\mathrm{ep}}$. 
Each of those charge carriers could get trapped in each step of the drift as described earlier. 
In the fit, the seed for the random numbers was fixed. 
Thus, the same random numbers were used for each set of parameters
and, therefore, each specific set of parameters always resulted in the same simulated energies. 
It was tested and verified that the large number of charge carriers ensured
that the simulated energies of the events were basically independent, $\mathcal{O}$(1\,keV), of the chosen seed.

The likelihood, $\mathcal{L}$, that a set of parameters represents the data was calculated using simulated energies, $E_{\mu,j,\mathrm{sim}}^{0/19}$, 
in the core and segment~19 for events spawned at all measurement positions of the scan RSF.
The data are the most likely observed alpha energies, $E_{\mu,j,\mathrm{obs}}^{0/19}$, 
as shown in Fig.~\ref{fig:rdep}a for each position of the scan RSF for the core and segment~19.

The simulation does not take certain systematic effects into account\footnote{E.g.\,possible inhomogeneities of the crystal and passivation layer, 
varying temperatures $\mathcal{O}$(5\,K) in GALATEA, details of the impurity profile, charge cloud effects, no (simulated) trapping along the widening of the borehole.}. 
Therefore, the simulated core and segment~19 energies were associated with an estimated uncertainty of $\sigma_{\mathrm{sim}} = 100\,$keV.
The individual likelihood, $l_j^{0/19}$, for each core and segment~19 energy at each radial position is calculated as 
\begin{equation}
    l_j^{0/19} = \mathcal{N}(E_{\mu,j,\mathrm{sim}}^{0/19}, \sigma_{\mathrm{sim}})(E_{\mu,j,\mathrm{obs}}^{0/19})\hspace{10pt},
\end{equation}
where $\mathcal{N}(E_{\mu,j,\mathrm{sim}}^{0/19}, \sigma_{\mathrm{sim}})$ is a Normal distribution with mean $E_{\mu,j,\mathrm{sim}}^{0/19}$
and a standard deviation of $\sigma_{\mathrm{sim}}$.
The overall $\mathcal{L}$ is given as the product over all positions:
\begin{equation}
    \mathcal{L} = \prod_j l_j^{0}\cdot l_j^{\mathrm{19}}\hspace{10pt}.
\end{equation}

\begin{figure}[htbp]
    \centering
    \includegraphics[width=\columnwidth]{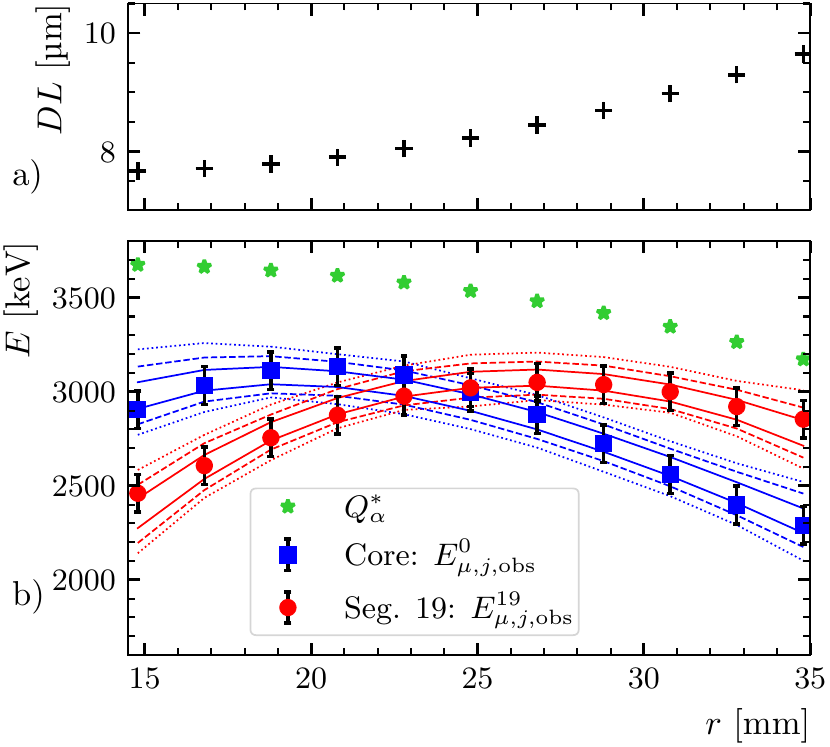}
    \caption{
    a) Dead-layer thickness as given by the fit to the data of the scan RSF, see Fig.~\ref{fig:rdep}a.
    b) Deposited energy $Q_{\alpha}^*$ corresponding to the depth of the dead layer and simulation results in form of 
    (1\,$\sigma$, 2\,$\sigma$, 3\,$\sigma$) posterior-predictive credibility intervals (solid, dashed and dotted lines) for 
    $E_{\mu,j,\mathrm{sim}}^{0}$ and $E_{\mu,j,\mathrm{sim}}^{19}$ compared to 
    $E_{\mu,j,\mathrm{obs}}^{0}$ and $E_{\mu,j,\mathrm{obs}}^{19}$ of the scan RSF.
    The error bars do not show the uncertainty on the data points but $\sigma_{\mathrm{sim}} =$\,100\,keV.
    }
    \label{fig:simrdep}
\end{figure}

\begin{figure}[htbp]
    \centering
    \includegraphics[width=\columnwidth]{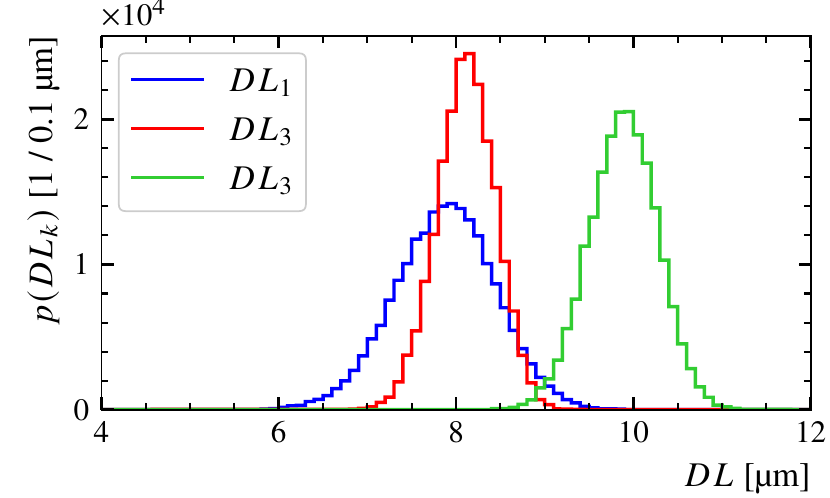}
    \caption{Marginalised posterior distribution, $p(DL_k)$, of the thickness of the dead layer at $r =$\,12.8\,mm ($DL_1$), $r =$\,23.0\,mm ($DL_2$) and $r =$\,34.8\,mm ($DL_3$).}
    \label{fig:sim_marginalized_posterior_dists_dead_layer}
\end{figure}

\begin{figure}[htbp]
    \centering
    \includegraphics[width=\columnwidth]{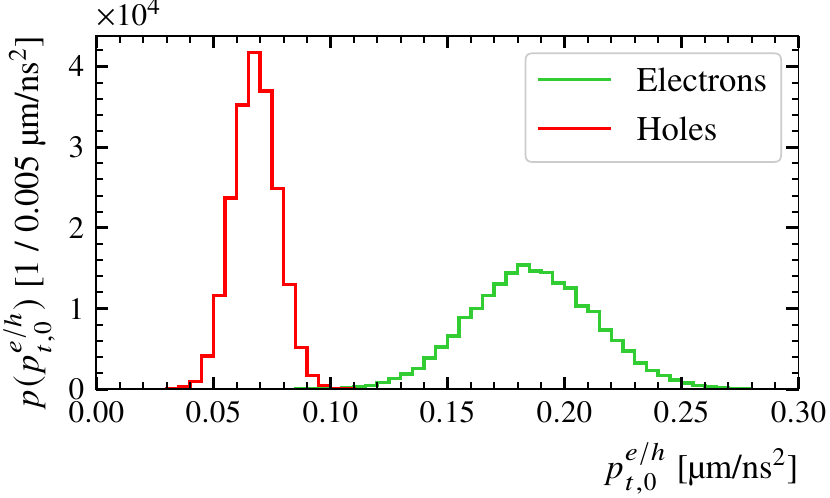}
    \caption{Marginalised posterior distribution, $p(p_{t,0}^{e/h})$, of the two parameters describing the charge-trapping probability.}
    \label{fig:sim_marginalized_posterior_dists_trapping_parameters}
\end{figure}

The result of the fit is shown in Fig.~\ref{fig:simrdep}. 
The simulation follows the trend of the data very well.
The marginalised posterior distributions of the parameters $DL_1$, $DL_2$ and $DL_3$ are shown in Fig.~\ref{fig:sim_marginalized_posterior_dists_dead_layer} and of the parameters $p_{t,0}^e$ and $p_{t,0}^h$ in Fig.~\ref{fig:sim_marginalized_posterior_dists_trapping_parameters}.
The fit suggests that the dead-layer thickness increases towards the mantle of the detector reducing the amount of energy deposited in the active volume. 
It also suggests that the trapping probability is about a factor of three higher for electrons than for holes.

\begin{figure*}[htbp]
    \centering
    \includegraphics[width=\columnwidth]{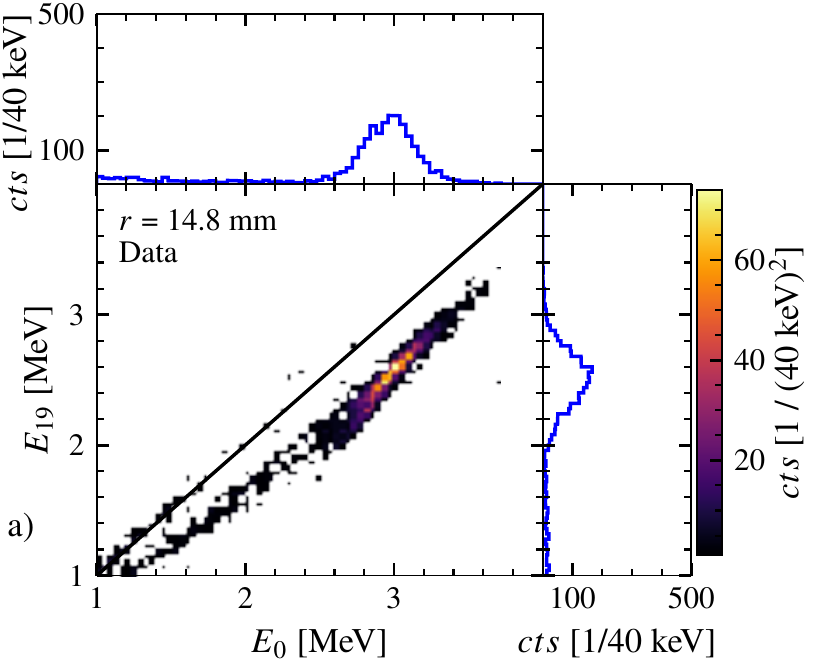}
    \includegraphics[width=\columnwidth]{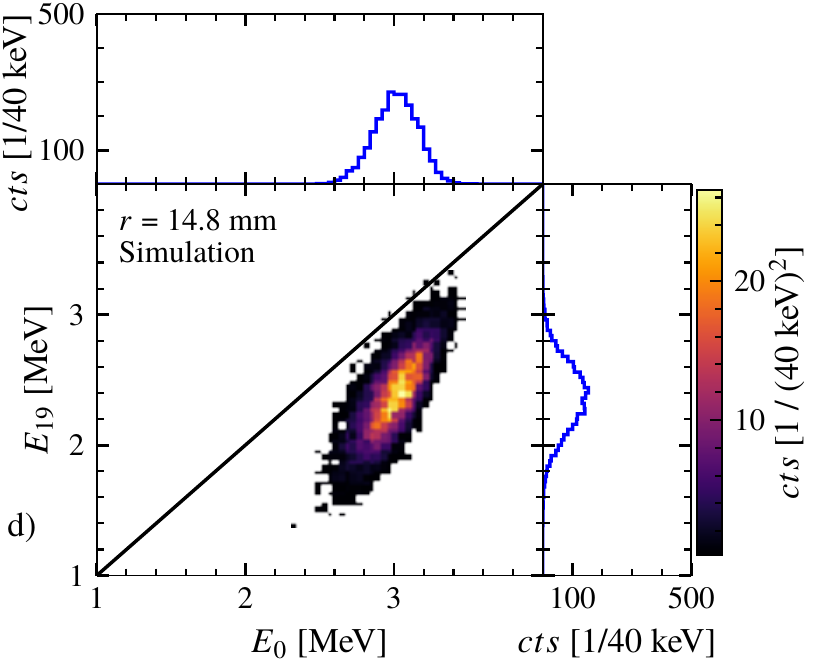}
    \includegraphics[width=\columnwidth]{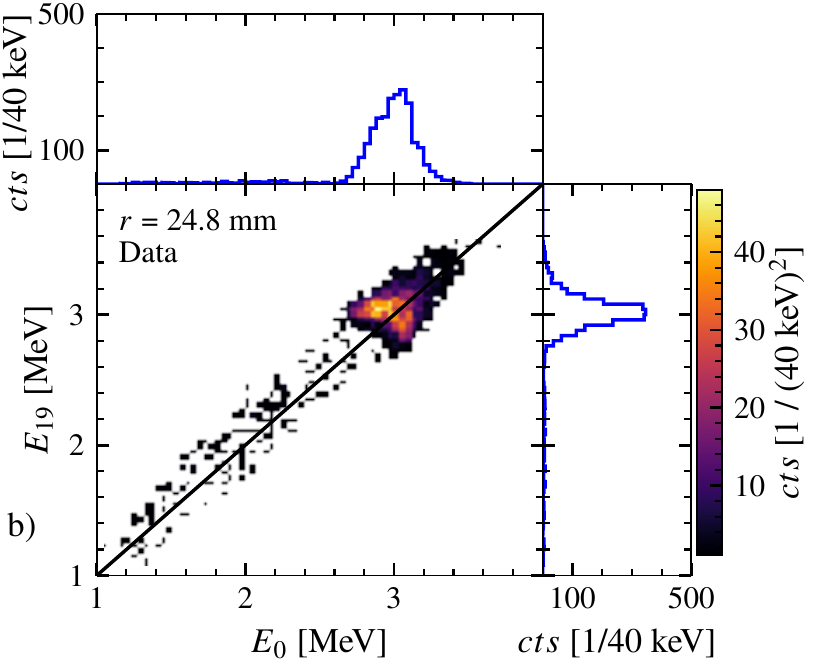}
    \includegraphics[width=\columnwidth]{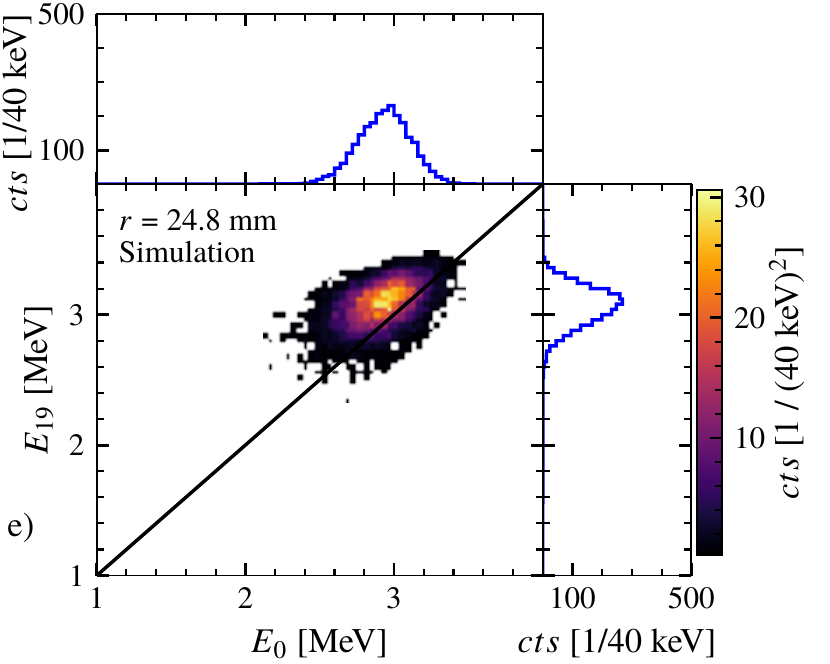}
    \includegraphics[width=\columnwidth]{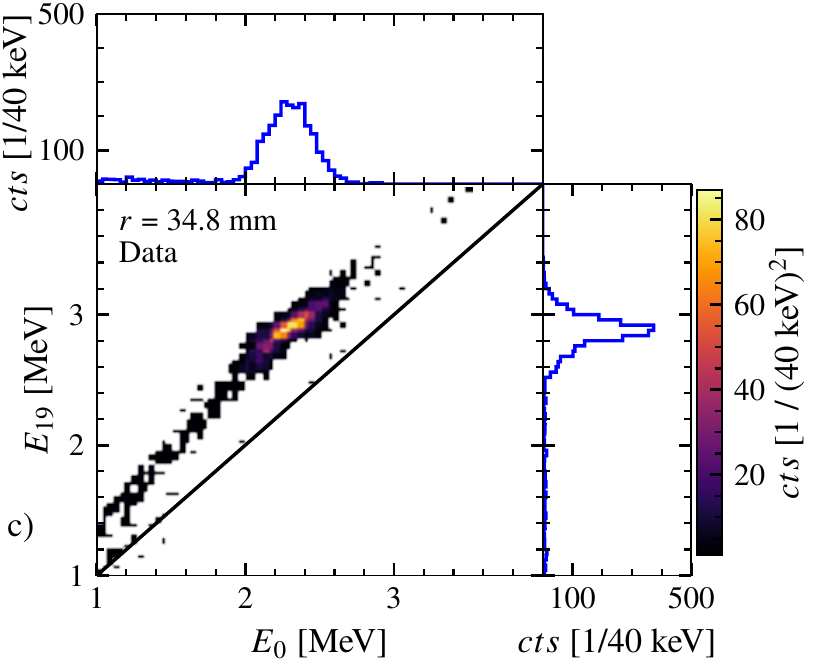}
    \includegraphics[width=\columnwidth]{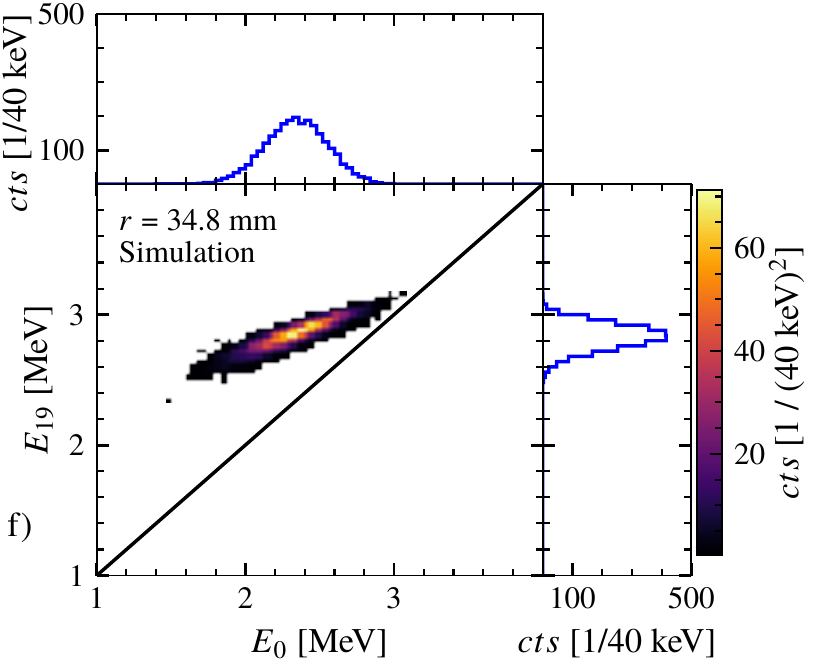}
    \caption{Correlation between core and segment~19 energies and their marginalisations next to them.
    On the left (a, b, c): Alpha events selected as described in Sec.~\ref{sec:selection} of three measurements from RSF. 
    On the right (d, e, f): Simulated alpha events spawned at the position of the respective measurement shown on the left. 
    The histograms obtained from the simulations are scaled to the respective histogram on the left.}
    \label{fig:Es19overEcore_data_vs_sim}
\end{figure*}

The parameters of the global mode of the fit were used to simulate 10000 events at each radial position of the scan RSF.
Each event was simulated with $N_{\mathrm{ep}} = 30$ and no seed was set.
The distributions of $E_{\mathrm{sim}}^{0}$ and $E_{\mathrm{sim}}^{19}$ and measured $E_{\mathrm{data}}^{0}$ and $E_{\mathrm{data}}^{19}$ are shown in Fig.~\ref{fig:Es19overEcore_data_vs_sim} for the three radial positions 14.8\,mm, 24.8\,mm and 34.8\,mm.
Different values for $N_{\mathrm{ep}}$ were tried and $N_{\mathrm{ep}} = 30$ was found to produce as broad energy distribution like seen in the data due to statistical fluctuations.
The relatively low value of $N_{\mathrm{ep}} = 30$ resulting in reasonably broad distributions could mean that the trapping mechanism actually happens in some coherent way, 
such that always a group of electrons or holes gets trapped during the drift. 
Another explanation might be that there are certain locations at the surface where charge carriers are more likely to 
get trapped such that, again, a group of electrons or holes is trapped at the same time or, respectively, at the same location.

\begin{figure}[htbp]
    \centering
    \includegraphics[width=\columnwidth]{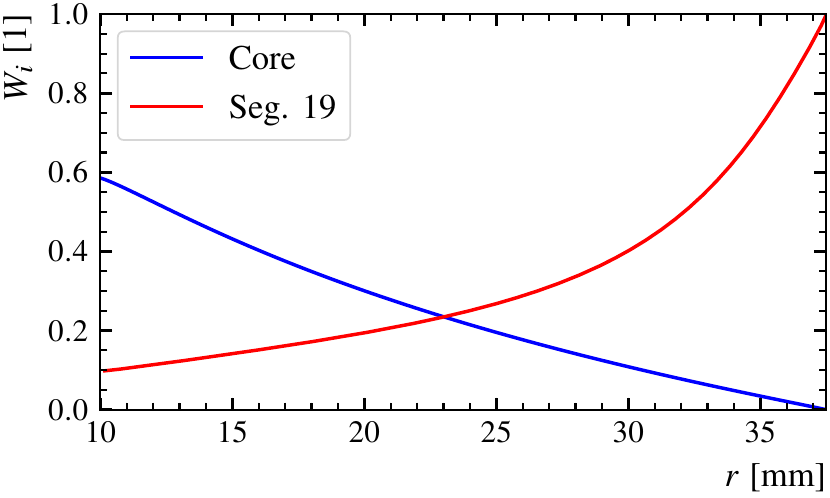}
    \caption{Weighting potentials of the core and segment~19 along $r$ at the top surface of the detector at $\varphi = 135.6^{\circ}$ (fast axis) and $z = 70$\,mm as calculated with SSD.
    }
    \label{fig:wps_at_top_surface}
\end{figure}

Even though the model was not tuned to predict correlations,
correlations similar to the observed correlations are produced.
For events at lower and larger radii, see Fig.~\ref{fig:Es19overEcore_data_vs_sim}a and Fig.~\ref{fig:Es19overEcore_data_vs_sim}c, 
the measured 2D-distributions are narrow and stretched.
This is also predicted by the simulation, see Fig.~\ref{fig:Es19overEcore_data_vs_sim}d and Fig.~\ref{fig:Es19overEcore_data_vs_sim}f. 
It can be explained by the shapes of the weighting potentials of the contacts, see Fig.~\ref{fig:wps_at_top_surface}.
The gradient of the weighting potential increases close to the respective contact.
For events at larger radii, the holes have only a short drift path
towards segment~19 and, thus, less holes are trapped during the drift. 
In addition, trapped holes still contribute significantly to the segment~19 signal as they are close to the contact. 
The electrons, however, are trapped at lower radii where the gradient of the 
weighting potential of segment~19 is basically zero.
Thus, the exact location of the trapping of electrons does not influence the signal of segment~19 
after the electrons moved inwards below a certain radial position, 
where the weighting potential of segment~19 is almost constant. 
Therefore, for events at small (large) radii the spread of $E_{19}$ ($E_{0}$) for a given $E_{0}$ ($E_{19}$) is smaller than the spread of $E_{0}$ ($E_{19}$) for a given $E_{19}$ ($E_{0}$).

At medium radii, see Fig.~\ref{fig:Es19overEcore_data_vs_sim}b, a half-moon shaped distribution is observed.
This is not reproduced by the simulation, see Fig.~\ref{fig:Es19overEcore_data_vs_sim}e.
However, this is probably because the spread of the beam-spot was not taken into account in the simulation.
All simulated events were spawned at the center of the beam-spot.
Taking the beam-spot into account, see Fig.~\ref{fig:geant4_top_beamspot},
would result in a superposition of the distributions for the different $r$ in the beam-spot. 

The simulation does not include the low-energy tails in the data which are
due to collimation effects, see Fig.~\ref{fig:geant4_energy_spectrum}.
It produces normally shaped distributions for $E_{0}$ and $E_{19}$, see Figs.~\ref{fig:Es19overEcore_data_vs_sim}d-f.
It is, however, interesting that the low-energy alphas seem to be affected by charge trapping
in the same way as the high-energy alphas. The trapping mechanisms seem to
depend only on the location and not on the total energy deposited.

\section{Effect of metalisation}
\label{sec:metal}

In previous studies~\cite{Abt:2016trw}, a strong dependence of the rise time of the pulses on $\varphi$ 
was observed for both alpha and gamma induced events in segment~19. 
For these studies, segment~19 was only metalised in a small area where the read-out cable was, and is, connected to the segment at $\varphi_{\mathrm{ro}}=135^{\circ}$ over a range of about 20$^{\circ}$. 
Since then, the detector was reprocessed and segment~19 became fully metalised.

\begin{figure}[htbp]
    \centering
    \includegraphics[width=\columnwidth]{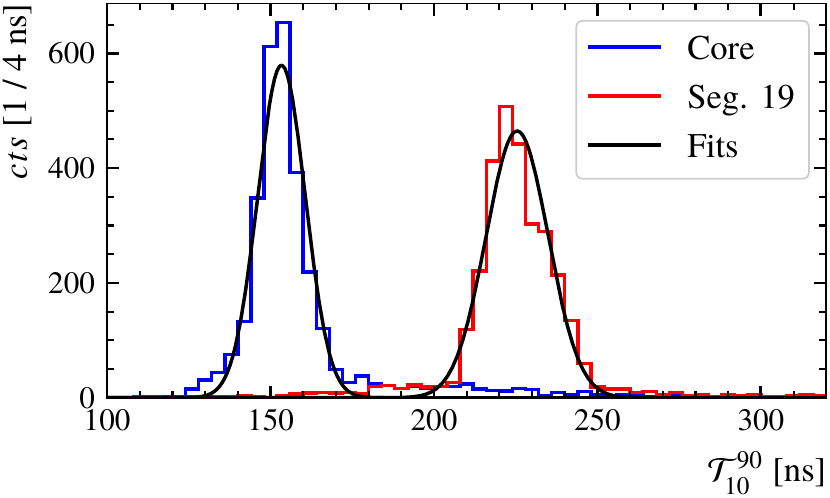}
    \caption{Distribution of the rise time $\mathcal{T}_{10}^{90}$ for the core and segment~19
    of the measurement at $\varphi=262.9^{\circ}$ from the scan AST. 
    Also shown are scaled Normal distributions fitted to the histograms.}
    \label{fig:rs1090_dists_for_alpha_events}
\end{figure}

\begin{figure*}[htbp]
    \centering
    \includegraphics[width=\textwidth]{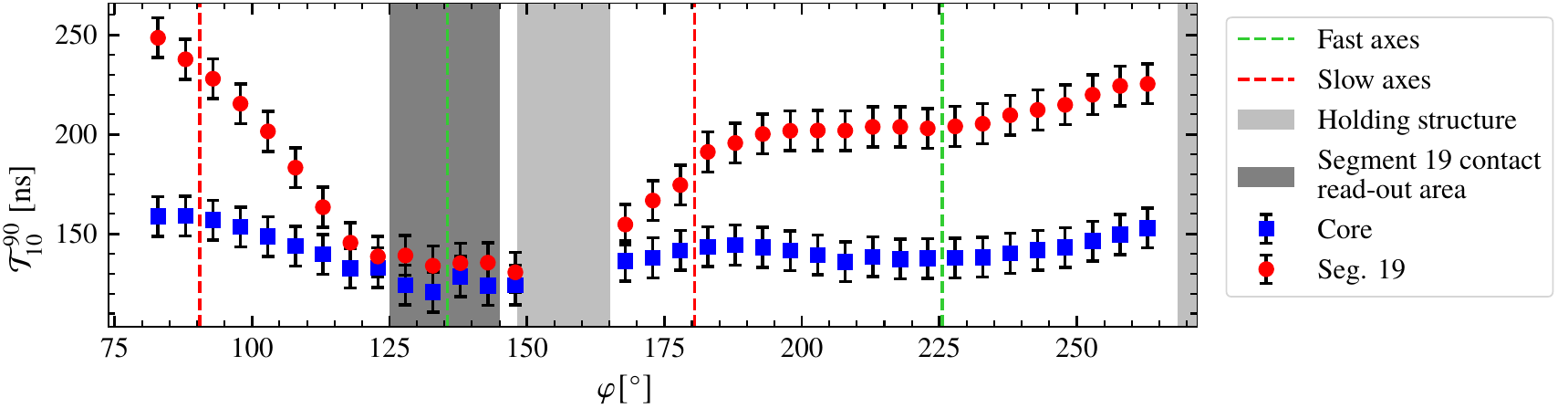}
    \caption{Fitted means of the distribution of $\mathcal{T}_{10}^{90}$ for the core and segment~19
    of alpha measurements from AST. 
    The error bars indicate systematic uncertainties due to the slightly fluctuating temperature of the detector in GALATEA~\cite{Hauertmann2017:MasterThesis}.
    }
    \label{fig:rs1090_mean_vs_phi}
\end{figure*}

The rise time, $\mathcal{T}_{10}^{90}$, defined as the time in which the pulse rises from 10\% to 90\% of its maximum amplitude, is used for the rise-time studies presented here.
The distribution of $\mathcal{T}_{10}^{90}$ of the core and segment~19 pulses of alpha induced events is shown in Fig.~\ref{fig:rs1090_dists_for_alpha_events} for one position of the scan AST.
The histograms were fitted with scaled Normal distributions for each position.
The fitted means of these distributions as a function of $\varphi$ are shown in Fig.~\ref{fig:rs1090_mean_vs_phi} for both the core and segment~19.

The core pulses show the expected modulation due to the crystal axes and 
a small dependence on $\Delta\varphi_j = |\varphi_{j} - \varphi_{\mathrm{ro}}|$, where $\varphi_{j}$
is the azimuthal position of the measurement $j$.
This is quantified by comparing the rise times at two positions close to different fast axes of the crystal.
At positions close to the fast axis at $\varphi \approx 225^{\circ}$, the rise time $\mathcal{T}_{10}^{90}$ is about 15\,ns longer than at positions close to the fast axis near near $\varphi_{\mathrm{ro}}$.

For segment~19, this effect is more pronounced and the influence of the crystal axes becomes less observable. 
Here, $\mathcal{T}_{10}^{90}$ is about 205\,ns at $\varphi \approx 225^{\circ}$ and only about 135\,ns at $\varphi_{\mathrm{ro}}$. 
This difference of about 70\,ns is, however, strongly reduced compared to the previous measurements~\cite{Abt:2016trw}, 
where a variation of the rise times of about 730\,ns in segment~19 
and about 50\,ns in the core was observed. 

The dependence of $\mathcal{T}_{10}^{90}$ on $\Delta\varphi_j$ is not only present for 
alpha induced events on the top surface but also for events induced on the top by
59.5\,keV gammas emitted from the Americium source~\cite{Hauertmann2021:PhDThesis}.

\begin{figure*}[htbp]
    \centering
    \includegraphics[width=\textwidth]{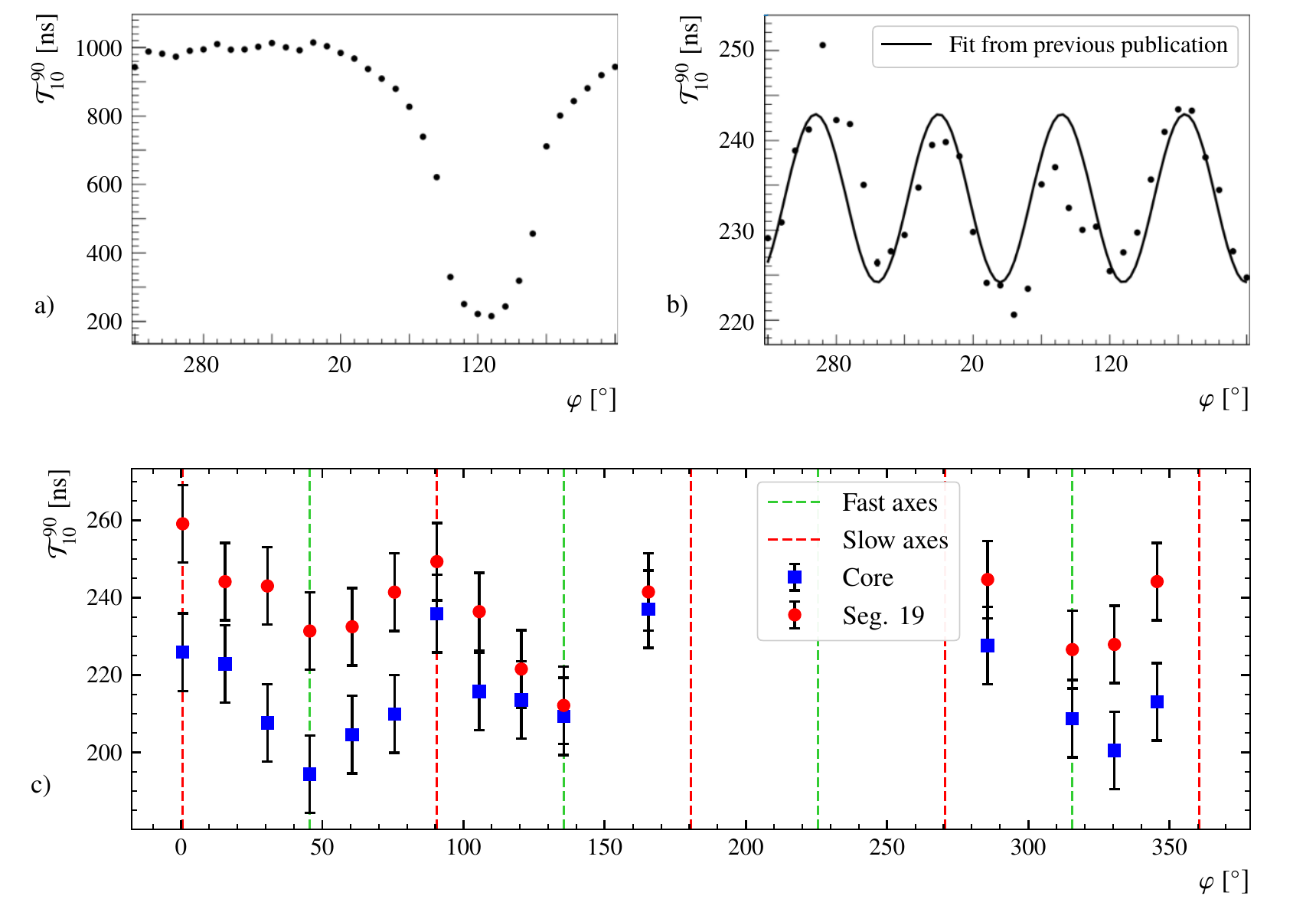}
    \caption{ 
        Averaged rise times $\mathcal{T}_{10}^{90}$ of events induced by collimated 121.8\,keV gammas from $^{152}$Eu entering segment~19 from the side for a) segment~19 and b) the core,
        adapted from \cite{PhD:Lenz2010,Abt:2016trw}. The $\varphi$-axes were modified to match the coordinate system chosen for this paper.
        c): Rise times $\mathcal{T}_{10}^{90}$ determined from the superpulses formed from the gamma events 
        induced by 59.5~keV gammas from the $^{241}$Am source from the rotational side scan ASS.
        The error bars indicate systematic uncertainties due to the slightly fluctuating temperature of the detector in GALATEA~\cite{Hauertmann2017:MasterThesis}.
    }
    \label{fig:rs1090_side_gammas_old_vs_now}
\end{figure*}

Prior to full metalisation, not only the rise times of events induced from the top were, 
but also events induced from 121.8\,keV gammas\footnote{Alpha events could not be induced on the side of segment~19 because it was, and is, covered with Kapton tape.}, entering  
segment~19 from the side of the detector from a $^{152}$Eu, were affected.
The values of $\mathcal{T}_{10}^{90}$ for segment~19 showed a very strong dependence on $\Delta\varphi_j$~\cite{PhD:Lenz2010,Abt:2016trw}
as shown in Fig.~\ref{fig:rs1090_side_gammas_old_vs_now}a. 
No such effect was observed for core pulses, see Fig.~\ref{fig:rs1090_side_gammas_old_vs_now}b. 
After the full metalisation, only a very small dependence of $\mathcal{T}_{10}^{90}$ on $\Delta\varphi_j$ is observable as shown 
in Fig.~\ref{fig:rs1090_side_gammas_old_vs_now}c where the rise times of superpulses\footnote{Superpulses are the averaged pulses of all selected gamma events of a measurement. They are used at these low energies
because the electronic noise is too large to determine $\mathcal{T}_{10}^{90}$ for individual pulses.}~\cite{Hauertmann2021:PhDThesis}
of 59.5\,keV gamma events from the scan ASS are depicted. 
In contrast to the previous measurements,
the modulation reflecting the crystal axes is now also visible in segment~19. 
This indicates that details of contacting and metalisation can influence pulse shapes significantly.
Such considerations could help to design germanium detectors such that events close to passivated surfaces could be easily identified.
One idea is to leave a gap between the passivation and the beginning of the metalisation.

\section{Summary and outlook}

Charge trapping was observed in alpha events on the passivated surface of the end-plate of an n-type true-coaxial segmented HPGe detector.
Such events occur with energies in the region around 2\,MeV where the signal for $0\nu\beta\beta$ decay in $^{76}$Ge would appear.
The trapping of holes and electrons was clearly identified by truncated mirror pulses observed on the segmented mantle of the detector.
This identification confirmed that alpha events can be selected effectively by analysing the tail slope of the core pulses.
Thus, the background due to alphas on the passivated surfaces can be efficiently suppressed for detectors as used 
in the first phase of the LEGEND experiment currently under construction at the Gran Sasso underground laboratory.

Radial scans of the detector top surface showed that charge trapping is dominated by hole trapping for events 
at small radii and by electron trapping for events at large radii as expected from the length of the respective drift paths.
The trapping probability per \textmu m/ns$^2$ was found to be three times larger for electrons than for holes
for drifts directly underneath the passivated top surface.
A dependence of the probability of charge trapping on the crystal axes has been observed for the first time in GALATEA.
Near a fast axis, $\langle100\rangle$, less net charge trapping is observed than near a slow axis, $\langle110\rangle$.
This can be understood as a consequence of higher mobilities along the fast axis.

Charge trapping was simulated with the open-source software package \emph{SolidStateDetectors.jl}.
The radial dependence of the probability of charge trapping can be described by a model including a radius-dependent dead layer, 
a modulation of the charge-carrier mobility and probabilistic charge trapping of electrons and holes.
In the future, the model can be extended to take crystal axes effects into account in order to describe the azimuthal dependence. 
In addition, new versions of \emph{SolidStateDetectors.jl} can take charge-cloud effects like diffusion and self-repulsion into account.
Furthermore, the non-zero tail slope, used to identify surface alpha events, could be added to the 
simulation by introducing a probabilistic release of trapped charge carriers in each step similar to the trapping probability,
or by letting the trapped charge carriers still drift with a very low velocity.

The influence of the segment metalisation on the pulse shapes of events close to the passivated top surface of the detector was studied.
In previous studies, very long pulses were observed close to the top for a metalisation scheme where only 
small areas were metalised. 
For a full metalisation, normal (faster) pulses and a much weaker 
dependence of the rise times on the event position with respect to the position of the read-out contact were observed.
However, a small $\varphi$ dependence beyond the crystal axes modulation was observed even for full metalisation.
Metalisation schemes could be used to influence the pulses of surface events to facilitate easier identification of such events.

\bibliographystyle{spphys} 

\bibliography{main.bib}

\end{document}